\documentclass[11pt]{article}

\usepackage[T1]{fontenc}

\usepackage{tikz}

\usepackage{array}
\usepackage{amsfonts,dsfont}
\usepackage{amsmath}
\usepackage{amssymb}
\RequirePackage[colorlinks,citecolor=blue,urlcolor=blue,linkcolor=blue]{hyperref}
\usepackage{graphicx,xspace,colortbl}
\usepackage{amsmath,amsthm,amsfonts,latexsym,amssymb}
\usepackage{color}
\usepackage{fancybox}
\usepackage{subfig}
\usepackage{pdfsync}
\usepackage{multirow}
\hypersetup{
colorlinks = true,
citecolor=blue,
urlcolor=blue,
linkcolor=blue,
pdfauthor = {Zied Ben-Salah, Helene Guerin, Manuel Morales, Hassan Omidi},
pdfkeywords = {Drawdowns, depletion, Levy insurance risk process, spectrally negative, scale function, risk theory},
pdftitle = {On the Depletion Problem for an Insurance Risk Process: New Non-ruin Quantities in Collective Risk Theory},
pdfpagemode = UseNone
}

\newtheorem{thm}{Theorem}

\newtheorem{pro}{Proposition}

\newtheorem{rem}{Remark}

    \def\p{{\mathbb P}}
    \def\e{{\mathbb E}}
    \def\beq{\begin{eqnarray}}
    \def\eeq{\end{eqnarray}}
    \def\beqq{\begin{eqnarray*}}
    \def\eeqq{\end{eqnarray*}}

\newcommand{\PAR}[1]{{\left(#1\right)}} 
\newcommand{\SBRA}[1]{{\left[#1\right]}} 
\newcommand{\BRA}[1]{{\left\{#1\right\}}} 
\newcommand{\ind}{\mathds {1}}

\title {On the Depletion Problem for an Insurance Risk Process: New Non-ruin Quantities in Collective Risk Theory\footnote{The authors acknowledge the initial input from Prof. Erhan Bayraktar of the University of Michigan that instigated this work. The authors would also like to thank the CNRS, its UMI 3457 and the CRM for providing the research infrastructure. This research was funded in part by the Natural Sciences and Engineering Research Council of Canada (NSERC)}}

\author{Zied Ben-Salah \footnote{University of Montreal, CANADA. Email: bensalah@dms.umontreal.ca}\\
University of Montreal \and
H\'el\`ene Gu\'erin 
\footnote{IRMAR. UMR CNRS 6625.
University of Rennes 1.
Campus de Beaulieu,
35042 Rennes Cedex. FRANCE. Email: helene.guerin@univ-rennes1.fr}\\
University of Rennes 1 
\and
Manuel Morales \footnote{Department of Mathematics and Statistics. University of Montreal. CP. 6128 succ. centre-ville. Montreal, Quebec. H3C 3J7. CANADA. Email: morales@dms.umontreal.ca} \\
University of Montreal \and
Hassan Omidi Firouzi \footnote{Hassan Omidi Firouzi. Department of Mathematics and Statistics. University of Montreal. CP. 6128 succ. centre-ville. Montreal, Quebec. H3C 3J7. CANADA. Email: omidifh@dms.umontreal.ca}\\
University of Montreal 
}

\date{{\scriptsize First draft: October~9, 2013. This version: \today.}}

\begin{document}
\maketitle


\begin{abstract}
{ The field of risk theory has traditionally focused on ruin-related quantities. In particular, the so-called Expected Discounted Penalty Function \cite{Gerber98a} has been the object of a thorough study over the years. Although interesting in their own right, ruin related quantities do not seem to capture path-dependent properties of the reserve. In this article we aim at presenting the probabilistic properties of drawdowns and the speed at which an insurance reserve depletes as a consequence of the risk exposure of the company. These new quantities are not ruin related yet they capture important features of an insurance position and we believe it can lead to the design of a meaningful risk measures. Studying drawdowns and speed of depletion for L\' evy insurance risk processes represent a novel and challenging concept in insurance mathematics. 
In this paper, all these concepts are formally introduced in an insurance setting. Moreover, using recent results in fluctuation theory for L\'evy processes \cite{Pistorius}, we derive expressions for the distribution of several quantities related to the depletion problem. Of particular interest are the distribution of drawdowns and the Laplace transform for the speed of depletion. These expressions are given for some examples of L\'evy insurance risk processes for which they can be calculated, in particular for the classical Cramer-Lundberg model. 
}
\end{abstract}


\section{Introduction}

Traditionally, collective risk theory is mainly concerned with the ruin problem which is nicely encapsulated in the concept of Expected Discounted Penalty Function (EDPF) introduced in \cite{Gerber98a}. This so-called Gerber-Shiu function is a functional of the ruin time (i.e., the first time the reserve level of a firm becomes negative), the surplus prior to ruin, and the deficit at ruin. 
The EDPF has been extensively studied and generalized to various scenarios and there is now a wide range of models for which expressions of the EDPF are available. All of these models incorporate different levels of complexity into the picture. 

In particular, the so-called L\'evy insurance risk processes have been the object of much attention in the last decade, mainly because they nicely generalize the Cramer-Lundberg model while allowing to bring new insight into the field of ruin theory through the well-developed theory of fluctuations for such processes. Several families of L\'evy processes have been put forward as risk models and we now have a well-established literature on the subject. For a thorough discussion on the suitability of these processes as risk models we refer the reader to \cite{Garrido_Morales,Morales} and references therein.

As it turns out, the first-passage problem for L\'evy processes is well understood and recent results in this area have been applied to the ruin problem in order to gain interesting insight (see for instance \cite{BK2010, BM2010, KM}). In this paper, we focus yet again on L\'evy insurance risk processes because of the extensive set of tools available for this family of stochastic processes. Through concepts originally developed for the study of the first-passage time problem, we can now study questions that go beyond the ruin problem and that are connected to path-properties of the process that give a tell-telling picture of how depletion occurs. 

Quantities such as the speed of depletion and drawdowns have been studied in finance in connection to the concept of market crash \cite{Z_H}. Indeed, in finance one would be interested in knowing how fast and how frequent drawdowns of a certain size occur. In insurance, these questions have not been studied yet, despite the fact that these concepts are meaningful from an insurance risk management point of view. Clearly knowing how your insurance reserve is affected by drawdowns and how fast and frequent these are could be useful to devise risk management tools. These quantities provide a measure of riskiness that is not linked to the ruin event but rather to the depletion features of the reserve. However, this problem is technically challenging due to the jump nature of insurance models.    

The aim of this paper is two-fold. One one hand, we aim at introducing the problem of depletion into the theory of collective risk theory as a meaningful question from a risk management point of view. We formally define new non-ruin quantities within the classical risk theory framework and we discuss their main features and advantages over traditional ruin-related quantities. Indeed, it is interesting to notice that all of the available research focuses on ruin-related quantities which, by their very nature, fail to explain how an insurance reserve depletes over time. Thus, although ruin theory provides a good probabilistic picture of the problem of insolvency of an insurance reserve, it cannot explain other features that are equally representative of the riskiness of an insurance reserve such as its speed of depletion and the frequency of drawdowns. The ruin event is an object of concern over the long-run but a risk manager might also keep an eye on any series of particularly large drawdowns especially if they happen particularly fast. So concerning oneself with the ruin event overlooks other risky events that also have an impact on the solvency and financial planning of an insurance company. 

A second objective is to actually derive expressions for the distribution of several depletion-related random variables. As it turns out, recent results in the theory of fluctuations for one-sided L\'evy processes \cite{Pistorius} can be used to derive expressions for these depletion-related quantities. Key to the derivation of such expressions is the scale function of the process driving the insurance risk model. As we discuss, general and non-explicit expressions for the distribution of random variables in the depletion problem can only be simplified if a simple form for the scale function is available. Hence, we derive explicit expressions for the case of a classical Cramer-Lundberg model driven by a compound Poisson process with exponential jumps. Not surprisingly, this simple case has always yielded text-book examples of closed-form solutions in the risk theory literature. The problem of depletion is no exception and we present explicit expressions for the distribution of several depletion-related random variables in this case. We also provide a similar analysis for other examples of insurance risk models possessing simple scale functions, namely the gamma subordinator and the stable family of processes.

This paper is organized as follows. In Section \ref{sec:Depletion} we introduce a general model based on a L\'evy risk process for which we define the depletion problem and the notions of {\it drawdowns} and {\it speed of depletion} as well as related variables. Some preliminary results from the theory of fluctuations for L\'evy processes are given in Section \ref{sec:preliminary}. In Section \ref{sec:general} we study the problem of depletion for an insurance L\'evy risk process and we give general expressions for the distribution of depletion random variables of interest. Finally, in Section \ref{sec:examples}, we derive explicit expressions for all depletion-related quantities for a three examples of L\'evy insurance risk processes.



\section{The Depletion Problem for an Insurance Risk Model}
\label{sec:Depletion}


We consider a very general setup that generalizes the standard Cramer-Lundberg model. We consider in this paper an insurance risk process $X=(X_t)_{t\geqslant 0}$ starting at an initial surplus $x\geqslant 0$ with $X$
a spectrally negative L\'evy process. For technical reasons that will become clear later in the paper, we restrict ourselves to those processes having paths of unbounded variation or paths of bounded variation as well as a Lévy measure which is absolutely continuous with
respect to the Lebesgue measure. In order to avoid the case of trivial reflected processes we exclude processes with monotone paths.
As is customary, the symbols $\e_x$ and $\p_x$ will denote the expectation and the probability measure related to the process started at $x$, and if the process is started from zero
we will use simple notations $\e$ and $\p$. 

Notice that a such a model 
 contains all elements of a traditional risk model and encompasses, among others, the risk models studied in \cite{BM2010, Furrer98, Huzak, Morales}. Indeed, the constant rate premium is included as the drift of $X$, the so-called perturbation comes in as the Brownian component of $X$ and the pure aggregate claims is present as the jump part of $X$, which could be set as a compound Poisson or an infinite activity process. With this in mind,
we assume the process $X$ to have a positive drift such that $\e[X_1]>0$. Notice that in traditional ruin theory, this assumption responds to both, technical and practical reasons. Technically, it is needed in order to avoid the possibility that $X$ becomes negative almost surely whereas from a practical point of view it makes sense since it is common practice in insurance to work with loaded premiums. Indeed, it is standard to write the drift component within $X$ in terms of a safety loading. For instance, notice that we can recuperate the classical Cramer-Lundberg model if $X_t=c\,t-S_t$ where $c:=(1+\theta) \mathbb E [S_1]$ and $S$ is a compound Poisson process modeling aggregate claims. The drift $c$, with a positive safety loading $\theta>0$, is the collected premium rate. In the context of the depletion problem we do not need this condition. We keep it here for purely practical reasons as it is common practice to have insurance loaded premiums. 


One of the advantages of considering a general L\'evy risk model 
is that we can use the tools and methods of the fluctuation theory of L\'evy processes, allowing for a somewhat deeper understanding of the ruin problem but also of the depletion problem which can prove to yield just as interesting information about the riskiness of the reserve.

For a more extensive discussion on L\'evy risk models we refer to \cite{Garrido_Morales}. 
In this paper we will specialize this setting to three examples of L\'evy processes that have been studied in the literature in the context of the ruin problem.

One of the main objects of interest in ruin theory is the \emph{ruin time}, $\tau$, representing the first passage time of an insurance Lévy risk process $X$ below zero when $X_0=x$, i.e. 
\begin{equation}
\label{def:ruin}
\tau:=\inf\{ t > 0 \; : \; X_t < 0 \}, 
\end{equation}
where we set $\tau=+\infty$ if $X_t\geq0$ for all $t\geqslant 0$. 

In this paper, our main object of concern is the depletion problem that has two different random times as its main building blocks. In order to give a thorough definition of these concepts we need to introduce some notation.

We define the running infimum and the running supremum of a given L\'evy process $X$ by 
$${\underline{X}}_t:=\inf_{0\leqslant s\leqslant t}  X_s \qquad  \mbox{and} \qquad \overline{X}_t:=\sup_{0\leqslant s\leqslant t}  X_s\;.$$

Now we characterize the depletion problem for $X$. We first define the {\it drawdown} process $Y=(Y_t)_{t\geqslant 0}$, associated with a given risk process $X$, to be 
\begin{equation}\label{def:drawdown}
Y_t := \overline{X}_t- X_t \;, \qquad t\geqslant 0 \; .
\end{equation}
The first-passage time over a level $a>0$ of the drawdown process $Y$ is then defined to be
\begin{equation}
\label{def:drawdown_time}
\tau_a := \inf \{ t\geqslant 0 : Y_t>a\} \;.
\end{equation}
It is well-known that $\tau_a<\infty$ $\mathbb P$-almost surely (see \cite{AKP}, Theorem 1).
Just like the ruin time in (\ref{def:ruin}), this new random time in (\ref{def:drawdown_time}) contains relevant information on potentially risky behavior of the reserve. Their distribution can be used to measure the likeliness of path-related events that might have a negative impact on the financial health of the reserve. The random time $\tau_a$ records the time at which a drawdown in the reserve is larger than a, previously agreed upon, critical level $a$. An interesting set of associated tale-telling random variables can be built upon the random time (\ref{def:drawdown_time}). First, we need to define a process that will be useful in constructing meaningful non-ruin quantities. The last time before $t$ that $X$ reaches its running supremum, denoted by $\overline{G}_t$, is defined as 
\begin{equation}
\label{def:G_time}
\overline{G}_t := \sup \{ s\leq t : X_s ~\text{or}~ X_{s-} = \overline{X}_s\} \;.
\end{equation}

Thus the time $\tau_a$ of the critical drawdown of size $a$ along with the following quantities characterize the depletion problem for $X$: 

\begin{itemize}
 \item the last time the reserve was at its maximum level prior to critical drawdown, $\overline{G}_{\tau_a}$; 
 \item the {\it speed of depletion}, $\tau_a-\overline{G}_{\tau_a}$; 
 \item the maximum reserve level attained before critical drawdown is observed, $\overline{X}_{\tau_a}$; 
  \item the minimum reserve level prior to critical drawdown, $\underline{X}_{\tau_a}$; 
 \item the largest drawdown observed before critical drawdown of size $a$, $Y_{\tau_a-}$; 
 \item the overshoot of the critical drawdown over level $a$, $Y_{\tau_a}-a$. 
 \end{itemize}
 \begin{figure}[h!]\label{fi:general}
\begin{center}
\input{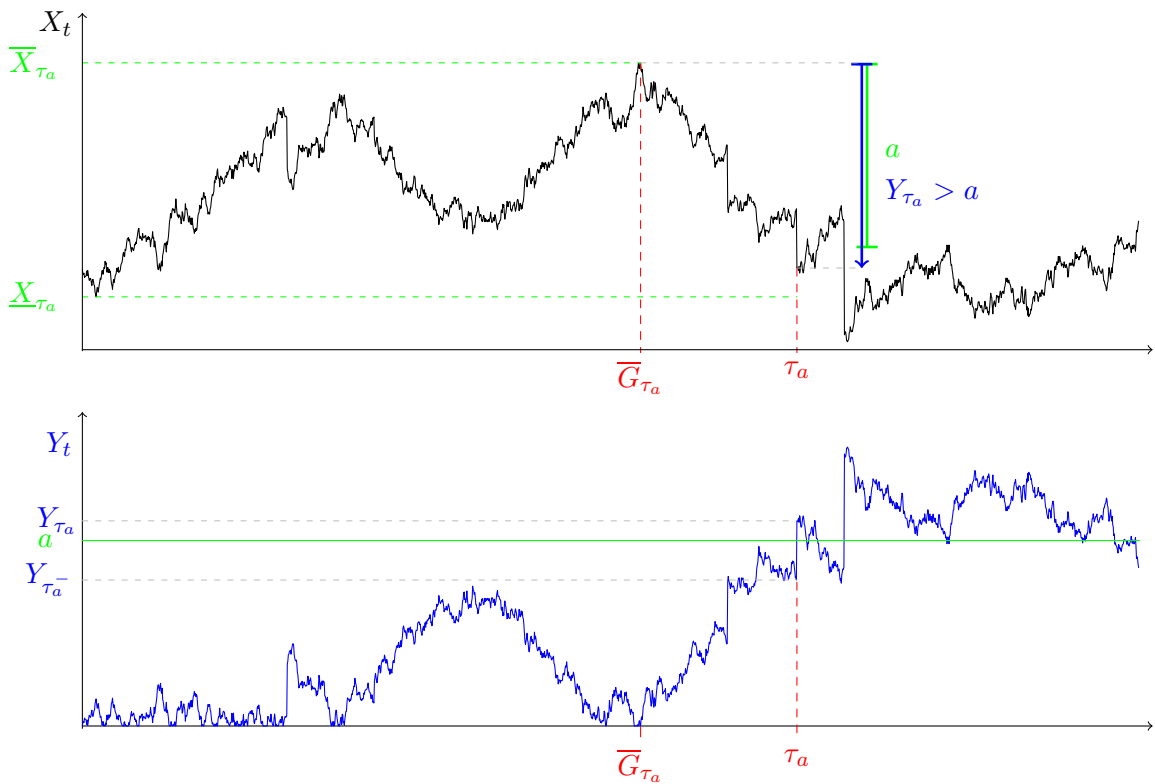}
\caption{A path of $X_t=10+t+2B_t-S_t$, the corresponding drawdown process $Y$, and their related depletion quantities, where $(B_t)_{t\geq 0}$ is a standard Brownian motion and $S$ is an independent compound Poisson process with L\'evy measure $\nu(dx)=e^{-2y}dy$.}
\end{center}
\end{figure} 

Clearly, these variables contain information on the how the insurance deserve depletes over time. All of these quantities encapsulate relevant knowledge about the critical drawdown event. A risk manager would be potentially interested in gaining information regarding the distribution of the time of the critical drawdown of size $a$, i.e. $\mathbb P_x (\tau_a \leqslant t)$. This gives information on how likely the reserve is to face a critical drawdown within a given time interval. Even more valuable information can be found in the distribution of the {\it speed of depletion}, this random variable indicates how fast critical drawdowns tend to occur. A drawdown is not as alarming if it happens over a long period than if it happens suddenly. Information about the distributions of the maximum and minimum reserve levels prior to critical drawdown and of the largest drawdown on record before critical drawdown sheds light on the structure of the depletion event. It is also interesting to know how large (or not) critical drawdowns tend to be, that is, by how much they overshoot the critical level $a$ when they occur. In fact, the level $a$ itself can be set by using the distribution of the overshoot. Since this distribution is a function of $a$ we can decide what a critical drawdown size is depending on how likely certain levels are. 

It is interesting to notice that there is a connection between ruin and depletion through the distribution of the minimum reserve level prior to critical drawdown. We will see that, if expressions are available, we can calculate the probability that ruin occurs before a critical drawdown of size $a$. 

In general, just like in the ruin problem, knowledge on the probabilistic properties of such quantities could be relevant in risk management applications. Although critical drawdowns do not spell immediate doom for the company as ruin does, a large enough drawdown might be a warning sign that a risk manager might want to take into account. This information coupled with knowledge on how fast these critical drawdowns happen could be used to design risk measures and or management policies that will ensure the solvency of the reserve. 

Once that these non-ruin quantities have been introduced, the aim of the paper is to derive expressions for the probability measure of these random variables associated with the depletion problem. This will be done in detail for three examples of L\'evy insurance risk processes. But before we need to introduce some preliminary results that are key to our analysis. 


\section{Drawdowns for Spectrally Negative L\'evy Processes}
\label{sec:preliminary}

In this subsection we introduce some notions and results that are needed in the rest of the paper. Let $X=(X_t)_{t\geqslant 0}$ be a spectrally negative Lévy process defined on a filtered probability space $(\Omega, \mathcal{F},(\mathcal{F}_t)_{t\geqslant 0}, \mathbb{P})$. We impose the same restrictions as in \cite{AKP}, i.e. $X$ has either paths of unbounded variation or
paths of bounded variation as well as a Lévy measure which is absolutely continuous with
respect to the Lebesgue measure. Further, we exclude processes with monotone paths.

Since $X$  has no positive jumps, the expectation $\mathbb{E}\SBRA{e^{s X_t}}$ exists for all $s\geqslant 0$ and it is given by $\mathbb{E}\SBRA{e^{s X_t}}= e^{ t \psi(s)}$ where $\psi(s)$ is of the form 
\begin{equation}
\label{khinchine}
\psi(s) = a \,s + \frac 12 \sigma^2 s^2 +\int_{0}^{\infty} (e^{-x\,s} -1 +s\, x \ind_\BRA{x<1}) \,\nu(dx)\;,
\end{equation}
where $a\in \mathbb{R}$,  $\sigma>0$ and $\nu$  is the L\' evy measure associated with the process $-X$ (for a thorough account on L\'evy process see \cite{Bertoin, Kyprianou}). 

For the right inverse of $\psi$, we shall write $\Phi$ on $[0,\infty)$. Formally, for each $q\geqslant 0$,
\begin{equation}
\label{right_inverse}
 \Phi(q):=\sup\{ s \geqslant 0: \psi(s)=q\} \; .
\end{equation}
Notice that since $X$ is a spectrally negative L\'evy process $X$, we have that $\Phi(q)> 0$ for $q > 0$ (see \cite{Kyprianou}).

It is well-known that, for every $q\geqslant 0$, there exists a function $W^{(q)}: \mathbb{R}\longrightarrow [0, \infty)$ such that $W^{(q)}(y)=0$ for all $y < 0$ satisfying
\begin{equation}
\label{scale1}
 \int_0^{\infty}e^{-\lambda y}W^{(q)}(y)dy=\frac{1}{\psi(\lambda)-q}, \qquad \lambda >\Phi(q).
\end{equation}
This is the so-called \emph{$q$-scale functions} $\{W^{(q)},\;q\geqslant 0\}$ of the process $X$ (see \cite{Kyprianou}) and it is a key notion in the analysis of drawdowns for spectrally negative L\'evy processes. Notice that for $q=0$, equation (\ref{scale1}) defines the so-called scale function and we simply write $W$. 

Before discussing drawdowns, we need to introduce additional functions related to the $q$-scale function.  Let $W_+'^{(q)}$ be the right derivative function of the $q$-scale. 
Following the notation in \cite{Pistorius}, we denote the  ratio of the right derivative of the $q$-scale function and the $q$-scale function at $a>0$ by
 \begin{equation}\label{Lambda}
 \lambda(a,q) := \frac{W_+'^{(q)}(a)}{W^{(q)}(a)}\;.
 \end{equation}
We can now define, for any $a>0$ and $p,q>0$, the mapping $F_{p,q,a}: \mathbb{R}_+\rightarrow \mathbb{R}_+$ 
 \begin{equation}\label{function}
 F_{p,q,a}(y) := \lambda(a,q) e^{-y\lambda(a,p)}\;.
 \end{equation}
Moreover, consider the q-resolvent measure  $R_a^{(q)}(dy)=\mathbb{E}\SBRA{\int_0^{\tau_a}e^{-qt}\ind_{Y_t\in dy}dt}$ of $Y$ killed upon first exit from $[0,a]$ which can be expressed in the following way (see \cite{pistorius04}, Theorem 1),
\begin{equation}\label{resolvent}
R_a^{(q)}(dy):= \left[ \lambda(a,q)^{-1} W^{(q)}(dy)-  W^{(q)}(y) dy\right]\;,\qquad y\in[0,a] \;,
\end{equation}
and the function
\begin{equation}\label{eq:delta}
\Delta^{(q)}(a)={\sigma^2\over 2}\SBRA{W'^{(q)}(a)-\lambda(a,q)^{-1}W''^{(q)}(a)}
\end{equation}
with $\Delta^{(q)}(a)=0$ when $\sigma=0$.
 
The functions in (\ref{Lambda}), (\ref{function}), (\ref{resolvent}) and (\ref{eq:delta}) will frequently appear throughout the paper. The following theorem will play a key role in our contribution. For a thorough discussion and a proof, we refer to \cite{Pistorius}.

 \begin{thm}\label{main}Consider a spectrally negative L\'evy process $X$ such that $X_0 = x \in \mathbb{R}$. Moreover, $X$ has paths of unbounded variation or has a L\'evy measure which is absolutely continuous with respect to the Lebesgue measure. Let further $Y$ be its associated drawdown process defined in (\ref{def:drawdown}). Let $\tau_a$ be the stopping time in (\ref{def:drawdown_time}) so we can define the following events, for a given $a>0$,  
\begin{equation}\label{A}
  A_0 = \{\underline{X}_{\tau_a}\geqslant u, \overline{X}_{\tau_a}\in dv, Y_{\tau_a-}\in dy, Y_{\tau_a}-a \in dh \}\quad \text{and}\quad A_c=\BRA{\underline X_{\tau_a}\geqslant u,\overline X_{\tau_a}\in dv,Y_{\tau_a}=a},
\end{equation} 
 \noindent where $u,v,y$ and $h$ satisfy
\begin{equation*}
u\leq x, ~y\in[0,a], ~v\geqslant x \vee (u+a) ~~\text{and}~~ h\in (0, v-u-a].
\end{equation*} 
Then, for any $q,r\geqslant 0$ the following identities hold true:
\begin{align}\label{identity}
\mathbb{E}_x \left[ e^{-q \tau_a -r\overline{G}_{\tau_a}} \ind_{A_0}\right]& = \frac{W^{(q+r)}((x-u)\wedge a)}{W^{(q+r)}(a)} F_{q+r, q,a}(v-(x\vee (u+a))) R_a^{(q)}(dy) \nu(a-y+dh) dv,\\
\nonumber\\
\mathbb{E}_x \left[ e^{-q \tau_a -r\overline{G}_{\tau_a}} \ind_{A_c}\right]& = \frac{W^{(q+r)}((x-u)\wedge a)}{W^{(q+r)}(a)} F_{q+r, q,a}(v-(x\vee (u+a))) \Delta^{(q)}(a)dv, \label{eq:Ac}
\end{align}
where $\ind$ is the standard indicator function, $\nu$ is the L\' evy measure of $X$ that appears in (\ref{khinchine}), $x\vee y = \max (x,y)$, $x\wedge y = \min (x,y)$ and $\overline{G}_t$ is the process defined in (\ref{def:G_time}).
  \end{thm}
We remark that the above theorem holds for spectrally negative L\'evy process having paths of unbounded variation or having a L\'evy measure which is absolutely continuous with respect to the Lebesgue measure. That's why we restrict ourselves to this type of processes. This is in no way restrictive since most of the risk insurance processes in the literature fall within this class, i.e. they are defined through a L\'evy density. 

We also remark that on the event $A_0$ defined  in (\ref{A}), the critical drawdown is performed by a jump of the L\'evy process $X$ while it is performed continuously on the event $A_c$.
These two events 
and the expectations in $(\ref{identity})$ and $(\ref{eq:Ac})$ in Theorem \ref{main} contain all information regarding the depletion problem. 
The aim of this paper, to provide explicit expressions for the distribution of these depletion-related random variables under relevant insurance models.

\section{Analysis of the Depletion Problem}
\label{sec:general}

In this section,
 we use the general setting described in Section \ref{sec:Depletion} where
 $X$ is a spectrally negative L\'evy process either with paths of unbounded variation or
paths of bounded variation with a Lévy measure absolutely continuous with
respect to the Lebesgue measure.
The main goal of this paper is then to study the depletion event as told by the quantities in Theorem \ref{main}. 
In principle, we can study through the expectations in (\ref{identity}) and \eqref{eq:Ac} the probability measure of all quantities involved as well as the Laplace transform of the speed of depletion. This can be accomplished by setting $q=r=0$ and/or integrating over a suitable set those expressions in (\ref{identity}) and \eqref{eq:Ac}. How explicit these expressions are will depend on the form of the $q$-scale function and the L\'evy measure of the model. Nonetheless, in this section we give some general results that bring insight into the problem.  

\medskip
It turns out, there is a link between the running infimum at time $\tau_a$, given by \eqref{def:drawdown_time}, and the ruin time $\tau$, given by \eqref{def:ruin}. We can easily deduce that $\tau_a\leq \tau$ a.s. on the event $\BRA{\underline{X}_{\tau_a}\geqslant 0}$, while $\BRA{\tau_a<\tau}$ implies  $\BRA{\underline{X}_{\tau_a}\geqslant  0}$.

Furthermore from the definition of these quantities, we can see that, when the initial surplus $x$ is strictly greater than $a$, then $\underline{X}_{\tau_a^-}>0$ and the hitting time of the critical drawdown is smaller than the ruin time, i.e. $\tau_a\leq \tau$ a.s. On the other hand, when $x<a$, ruin can occur before the critical drawdown.

We can now state a a result which makes a link between the ruin event and the depletion problem.

\begin{thm}\label{main1}
Consider an insurance risk process $(X_t)_{t\geqslant 0}$ with initial surplus $x\geqslant 0$ satisfying assumptions of Section \ref{sec:Depletion} and let $a>0$ be a fixed critical drawdown size. Then, 
\begin{equation}\label{ruin_before}
\mathbb{P}_x(\underline{X}_{\tau_a}<0)=1-\frac{W(x\wedge a)}{W(a)}+\frac{W(x\wedge a)}{W(a)}\int_{y\in[0,a]}\int_{h>0} \PAR{1-e^{-\lambda(a,0)h}} 
\nu(x\vee a-y+dh) R_a^{(0)}(dy),
\end{equation}
where $W$ is the scale function, $\nu$ the L\'evy measure of $X$ and $R_a^{(0)}$ is defined by \eqref{resolvent} with $q=0$. 
 \end{thm}

\begin{proof} 

We notice that $\mathbb{P}_x\PAR{\underline{X}_{\tau_a}<0}
=1-\mathbb{P}_x(\underline{X}_{\tau_a}\geqslant 0)$
and 
\begin{equation*}
\mathbb{P}_x(\underline{X}_{\tau_a}\geqslant 0)
= \mathbb{P}_x\PAR{\underline{X}_{\tau_a}\geqslant 0,\, Y_{\tau_a}>a}
+ \mathbb{P}_x\PAR{\underline{X}_{\tau_a}\geqslant 0,\,  Y_{\tau_a}=a}.
\end{equation*}
Then putting $q=r=u=0$ and integrating \eqref{identity} and \eqref{eq:Ac} with respect to $v\in[x\vee a,\infty)$, $y\in[0,a]$ and $h\in(0,v-a]$, we have
\begin{multline*}
\mathbb{P}_x(\underline{X}_{\tau_a}\geqslant 0)
=\frac{W(x\wedge a)}{W(a)}\left[ \int_{y\in[0,a]}\PAR{\int_{v\geqslant x\vee a}F_{0, 0,a}(v-x\vee a) 
\int_{h\in(0,v-a]} \nu(a-y+dh) dv}R_a^{(0)}(dy)\right.\\
\left.+\int_{v\geqslant x\vee a}F_{0, 0,a}(v-x\vee a) \Delta^{(0)}(a)dv\right].
\end{multline*}
Since $F_{0, 0,a}$ is defined by \eqref{function}, using Fubini's theorem, the previous expression gives
\begin{equation}\label{eq:probaruine}
\mathbb{P}_x(\underline{X}_{\tau_a}\geqslant 0)=\frac{W(x\wedge a)}{W(a)}\left[ e^{\lambda(a,0)x\vee a}\int_{y\in[0,a]}\int_{h>0} e^{-\lambda(a,0)(x\vee (h+a))} 
\nu(a-y+dh) R_a^{(0)}(dy)+ \Delta^{(0)}(a)\right].
\end{equation}

We notice that taking $u\rightarrow -\infty$ and integrating  \eqref{identity} and \eqref{eq:Ac} with respect to $v\in[x\vee a,\infty)$, $h\in(0,v-a]$ and $y\in[0,a]$ with $r=q=0$,
\begin{equation}\label{important}
\int_0^a R_a^{(0)}(dy) \int_0^{\infty}\nu(a-y+dh)+\Delta^{(0)}(a)= 1.
\end{equation}

Using this remark, we deduce that \eqref{eq:probaruine} gives for $x\leq a$
\begin{align*}
\mathbb{P}_x(\underline{X}_{\tau_a}\geqslant 0)&=\frac{W(x)}{W(a)}\left[ \int_{y\in[0,a]}\int_{h>0} e^{-\lambda(a,0) h} 
\nu(a-y+dh) R_a^{(0)}(dy)+ \Delta^{(0)}(a)\right]\\
&=\frac{W(x)}{W(a)}\left[1- \int_{y\in[0,a]}\int_{h>0} \PAR{1-e^{-\lambda(a,0) h} }
\nu(a-y+dh) R_a^{(0)}(dy))\right],
\end{align*}
and for $x>a$,
\begin{align*}
&\mathbb{P}_x(\underline{X}_{\tau_a}\geqslant 0)\\
&=\int_{y\in[0,a]}\int_{0}^{x-a}
\nu(a-y+dh) R_a^{(0)}(dy)+\int_{y\in[0,a]}\int_{x-a}^\infty e^{-\lambda(a,0)(h+a-x)} 
\nu(a-y+dh) R_a^{(0)}(dy)+ \Delta^{(0)}(a)\\
&=1-\int_{y\in[0,a]}\int_{x-a}^\infty \PAR{1-e^{-\lambda(a,0)(h+a-x)}} 
\nu(a-y+dh) R_a^{(0)}(dy)\\
&=1-\int_{y\in[0,a]}\int_{h>0} \PAR{1-e^{-\lambda(a,0)h}} 
\nu(x-y+dh) R_a^{(0)}(dy).
\end{align*}
The theorem is proved.

\end{proof}

Theorem \ref{main1} is of interest because $\mathbb{P}_x[\underline{X}_{\tau_a}<0]$ is in fact the probability of ruin occurring before a critical drawdown of size $a$, i.e. it is the probability that the reserve falls below the level zero during the interval $[0, \tau_a]$. 

In Section \ref{sec:generaldist}, we first give general expressions for the probability measures of depletion-related quantities. Since it might be of interest, from a risk management point of view, to study the depletion problem when ruin does not occur before the critical drawdown time, we also compute the distribution of depletion-related quantities on the event $\{\tau_a < \tau\}$. This is carried out in Section \ref{sec:ruindist}. 

\subsection{Distributions of depletion quantities}\label{sec:generaldist}

 
\begin{thm}\label{main2}
Consider an insurance risk process $(X_t)_{t\geqslant 0}$ with initial surplus $x\geqslant 0$ satisfying assumptions of Section \ref{sec:Depletion} and let $a>0$ be a fixed critical drawdown size. Then,
\begin{enumerate}
\item the probability distribution of the drawdown observed just before critical drawdown is the following,
\begin{equation}\label{eq:LawY-}
\mathbb{P}_x(Y_{\tau_a-} \in dy )= \PAR{R_a^{(0)}(dy)   \int_0^{\infty}\nu(a-y+dh)}\ind_{y\in (0,a]}+ \Delta^{(0)}(a) \delta_0(dy)\;,
\end{equation}
\item the probability distribution of the overshoot over the critical drawdown $Y_{\tau_a}-a$ is the following,
\begin{equation} \label{factorization17}
\mathbb{P}_x (Y_{\tau_a}-a \in dh) = \int_0^a  \nu(a-y+dh) R_a^{(0)}(dy)\ind_{h>0}+ \Delta^{(0)}(a) \delta_0(dh)\;,
\end{equation}
\item the maximum reserve level attained before critical a drawdown, $\overline{X}_{\tau_a}$, follows a translated exponential distribution, i.e.
\begin{equation*}
\mathbb{P}_x(\overline{X}_{\tau_a} \in dv)=  \lambda(a,0)e^{-\lambda(a,0)(v-x)}\ind_{v\geqslant x} dv\;.
\end{equation*}
\end{enumerate}
\end{thm}

\begin{proof}
In order to prove this result, we use Theorem $\ref{main}$ when $u \rightarrow -\infty$ with $r=q=0$. By integrating,
\begin{enumerate}

\item 
For $y\in[0,a)$, we have 
\begin{equation*}\label{factorization15}
\mathbb{E}_x[I_{\{Y_{\tau_a-} \in dy \}}]= R_a^{(0)}(dy) \int_x^{\infty}F_{0, 0,a}(v-x) dv  \int_0^{\infty}\nu(a-y+dh),
\end{equation*}
 and for $y=a$, 
$
\mathbb{P}\PAR{Y_{\tau_a-}=a}=\Delta_a^{(0)} \int_x^{\infty}F_{0, 0,a}(v-x) dv.
$

\item For $h>0$, we have
\begin{equation*}
\mathbb{E}_x[I_{\{Y_{\tau_a}-a \in dh \}}]=  \int_x^{\infty}F_{0, 0,a}(v-x) dv\int_0^a\nu(a-y+dh)R_a^{(0)}(dy),
\end{equation*}
and for $h=0$, $
\mathbb{P}\PAR{Y_{\tau_a}=a}= \int_x^{\infty}F_{0, 0,a}(v-x) dv\Delta^{(0)}(a).$

\item Finally, 
\begin{equation*}
\mathbb{P}_x(\overline{X}_{\tau_a} \in dv)=\mathbb{E}_x \left[ I_{\{\overline{X}_{\tau_a} \in dv\}}\right] = F_{0, 0,a}(v-x) dv\PAR{\int_0^a R_a^{(0)}(dy) \int_0^{\infty}\nu(a-y+dh)+\Delta^{(0)}(a)}.
\end{equation*}
Using \eqref{important} and (\ref{function}) yields the result.

\end{enumerate}
\end{proof}
From Theorem \ref{main2}, we notice that the distributions of $Y_{\tau_a^-}$ and of $Y_{\tau_a}-a$ do not depend on the initial surplus $x$ and whatever are the characteristics of the Lévy process $X$, the distribution of the maximum reserve level $\overline{X}_{\tau_a}$ attained before critical drawdown is always an exponential distribution shifted from the initial surplus $x$. This result is a typically extension of the same result where we study the distribution of the maximum reserve level $\overline{X}_{\mathtt{e}_q}$ attained before an exponentially distributed random time $\mathtt{e}_q$ with parameter $q$. 

Now we turn our attention to the random times $\tau_a$ and $\overline{G}_{\tau_a}$. We start by giving an interesting result concerning the speed of depletion $\tau_a -\overline{G}_{\tau_a}$. This is an immediate consequence from to Theorem \ref{main} that was not pointed out in \cite{Pistorius} and yet it is crucial to the actual evaluation of all components in the expression (\ref{identity}).  

\begin{pro}\label{lemma1}
Under the same assumptions and definitions of Theorem \ref{main}, the random variables $\overline{G}_{\tau_a}$ and $\tau_a -\overline{G}_{\tau_a}$ are independent.
\end{pro}
\begin{proof}
It can be easily verified that the statement in Theorem \ref{main} still holds under weaker conditions on $q$ and $r$. In fact, the result in (\ref{identity}) holds true for $q\geqslant 0$ and $q+r\geqslant0$ and not only for $q,r\geqslant 0$ as indicated in the original statement.  In fact, the conditions on $q$ and $r$ arise in the proof when we want to take the $q$ and  $q+r$ -scale functions into account in the expressions given in Theorem \ref{main}.  As the scale functions $W^{(q)}, W^{(q+r)}$ are just well defined  for $q, q+r\geqslant 0$.

By definition, $\overline{G}_{\tau_a}$ and $\tau_a-\overline{G}_{\tau_a}$ are positive $\mathbb P$-almost-surely finite random variables. It is well-known that, for $r, q\geqslant 0$, the bivariate Laplace transform $\mathbb{E}_x\left[e^{-r \overline{G}_{\tau_a} -q (\tau_a -\overline{G}_{\tau_a})}\right]$ characterizes the joint distribution of $\overline{G}_{\tau_a}$ and $\tau_a-\overline{G}_{\tau_a}$ (see for example \cite{feller}). 

Clearly,
\begin{equation*}
\mathbb{E}_x \left[ e^{-r \overline{G}_{\tau_a} -q (\tau_a -\overline{G}_{\tau_a})} \right] = \mathbb{E}_x \left[ e^{-q \tau_a -(r-q) \overline{G}_{\tau_a}} \right]\;,
\end{equation*}
and so an expression for the bivariate Laplace transform of $\overline{G}_{\tau_a}$ and $\tau_a-\overline{G}_{\tau_a}$ can be obtained through identities (\ref{identity}) and \eqref{eq:Ac} in Theorem \ref{main}.  Since $F_{r, q,a}(v)$ is the product of a function depending only of $r$ and a function depending only on $q$, Expressions  (\ref{identity}) and \eqref{eq:Ac} are also the product of a function depending only of $r$ and a function depending only on $q$ respectively, which conclude the proof.

\end{proof}

\begin{pro}\label{lemma_L}
Consider an insurance risk process $(X_t)_{t\geqslant 0}$ with initial surplus $x\geqslant 0$ satisfying the assumptions of Section \ref{sec:Depletion} and let $a>0$ be a fixed critical drawdown size. Then, for $q\geqslant 0$, $q+r\geqslant 0$, 
 the bivariate Laplace transform of $\tau_a$ and $\overline{G}_{\tau_a}$ is given by
\begin{equation}\label{bivariate_L}
\mathbb{E}_x \left[ e^{-q \tau_a -r\overline{G}_{\tau_a}}\right ] =\frac{\lambda(a,q)}{\lambda(a, q+r)}  \PAR{\int_{y\in[0,a]} \int_{h>0}  \nu(a-y+dh) R_a^{(q)}(dy)+ \Delta^{(q)}(a)}\;.
\end{equation}
\end{pro}

\begin{proof}Now, just like in the proof of Proposition \ref{lemma1}, we notice that the result of Theorem \ref{main} is still valid with $q\geq0$ and $r+q\geq0$. 
Taking $u \rightarrow -\infty$ and integrating \eqref{identity} and \eqref{eq:Ac} with respect to $v$, $h$ and $y$ in Theorem \ref{main}, we obtain 
\begin{equation*}
\mathbb{E}_x \left[ e^{-q \tau_a -r\overline{G}_{\tau_a}}\right ] =\int_{v\geqslant x}F_{q+r, q,a}(v-x) dv \PAR{ \int_{y\in[0,a]} \int_{h>0}  \nu(a-y+dh) R_a^{(q)}(dy)+ \Delta^{(q)}(a)}\;.
\end{equation*}
From the definition \eqref{function} of  $F_{q+r, q,a}(.)$, we have 
$\int_x^{\infty}F_{q+r, q,a}(v-x) dv = \frac{\lambda(a,q)}{\lambda(a, q+r)}$. Substituting this last equation yields,
\begin{equation*}
\mathbb{E}_x \left[ e^{-q \tau_a -r\overline{G}_{\tau_a}}\right ] =\frac{\lambda(a,q)}{\lambda(a, q+r)} \PAR{\int_{y\in[0,a]} \int_{h>0}  \nu(a-y+dh) R_a^{(q)}(dy)+ \Delta^{(q)}(a)}\;.
\end{equation*}
\end{proof}

\begin{rem} Putting $r=0$ in \eqref{bivariate_L}, the Laplace transform of $\tau_a$ is given by  
\begin{equation*}
\mathbb{E}_x\SBRA{e^{ -q \tau_a}} =  \int_{y\in[0,a]} \int_{h>0}  \nu(a-y+dh) R_a^{(q)}(dy)+ \Delta^{(q)}(a)\;.
\end{equation*}
Using \eqref{bivariate_L} with $q=0$ and \eqref{important} the Laplace transform of $\overline{G}_{\tau_a}$ is given by
\begin{equation}\label{Laplace_G_time}
\mathbb{E}_x\SBRA{e^{ -r \overline{G}_{\tau_a}}} =  \frac{\lambda(a,0)}{ \lambda(a,r)}\;.
\end{equation}
\end{rem}

In the following, we are going to provide an expression for the Laplace transform of the depletion random variable, $\tau_a -\overline{G}_{\tau_a}$.

\begin{thm}\label{Laplace_speed}
Consider an insurance risk process $(X_t)_{t\geqslant 0}$ with initial surplus $x\geqslant 0$  satisfying the assumptions of Section \ref{sec:Depletion} and let $a>0$ be a fixed critical drawdown size. Then, the Laplace transform of the speed of depletion $\tau_a -\overline{G}_{\tau_a}$ is given by 
\begin{equation*}
 \mathbb{E}_x\SBRA{e^{-q (\tau_a -\overline{G}_{\tau_a})}} = \frac{\lambda(a,q)}{\lambda(a,0) } \PAR{\int_{y\in[0,a]} \int_{h>0}  \nu(a-y+dh) R_a^{(q)}(dy)+\Delta^{(q)}(a)}\;.
 \end{equation*}
\end{thm}

\begin{proof}
By Proposition \ref{lemma1}, we know that $\overline{G}_{\tau_a}$ and  $\tau_a -\overline{G}_{\tau_a}$ are independent variables, then for $r, q\geqslant 0$
\begin{equation}\label{eq:L_s}
\mathbb{E}_x\SBRA{e^{-r \overline{G}_{\tau_a}}} \mathbb{E}_x\SBRA{e^{-q (\tau_a -\overline{G}_{\tau_a})}} =\mathbb{E}_x\SBRA{e^{-r \overline{G}_{\tau_a} -q (\tau_a -\overline{G}_{\tau_a})}} = \mathbb{E}_x\SBRA{e^{-q \tau_a -(r-q) \overline{G}_{\tau_a}}}\;.
\end{equation}


We can now find an expression for the right-end of equation (\ref{eq:L_s}) by setting $q^*=q$ and $r^*=r-q$ and using (\ref{bivariate_L}). In other words, since $q^*\geqslant 0$ and $q^*+r^*\geqslant 0$, we can then write 
\begin{multline*}
\mathbb{E}_x\SBRA{e^{-q \tau_a -(r-q) \overline{G}_{\tau_a}}} = \mathbb{E}_x\SBRA{e^{-q^* \tau_a -r^* \overline{G}_{\tau_a}}} \\
=  \frac{\lambda(a,q^*)}{\lambda(a, q^*+r^*)}  \PAR{\int_{y\in[0,a]} \int_{h>0}  \nu(a-y+dh) R_a^{(q^*)}(dy)+\Delta^{(q^*)}(a)}\;.
\end{multline*}
Substituting $q^*=q$ and $r^*=r-q$ into (\ref{inter}) and using equation (\ref{eq:L_s}) and equation (\ref{Laplace_G_time}) in Proposition \ref{lemma_L} yield the result. 
\end{proof}


\subsection{Distributions of depletion quantities in risk management} \label{sec:ruindist}

In this section we study expressions for the conditional distribution of depletion quantities discussed in Section \ref{sec:generaldist} given the event $\BRA{\underline{X}_{\tau_a}\geqslant 0}$. This set, guarantees that ruin does not occur before the critical drawdown time. In the following section, we use the notations $\mathbb{P}(.\, ;A)$ and  $\mathbb{E}\SBRA{.\, ;A}$  for $\mathbb{P}\PAR{.\cap A}$ and $\mathbb{E}\SBRA{.\, \ind_{A}}$ respectively.

\begin{pro}\label{conditional}
Consider an insurance risk process $(X_t)_{t\geqslant 0}$ satisfying assumptions of Section \ref{sec:Depletion} with initial surplus $x>0$ and let $a>0$ be a fixed critical drawdown size. Then,
\begin{enumerate}
\item the conditional distribution of the drawdown observed just before critical drawdown given the event $\BRA{\underline{X}_{\tau_a}\geqslant 0}$ is 
\begin{multline*}
\mathbb{P}_x\PAR{Y_{\tau_a-} \in dy|~ \underline{X}_{\tau_a}\geqslant 0 }\\
= \frac{R_a^{(0)}(dy)   e^{\lambda(a,0)x\vee a}\int_{h>0} e^{-\lambda(a,0)(x\vee (h+a))} 
\nu(a-y+dh) + \Delta^{(0)}(a) \delta_0(dy)}{1 - \int_{y\in[0,a]}\int_{h>0} \PAR{1-e^{-\lambda(a,0)h}} 
\nu(x\vee a-y+dh) R_a^{(0)}(dy)}\;,
\end{multline*}
\item the conditional distribution of the overshoot over the critical drawdown $Y_{\tau_a}-a$ given the event $\BRA{\underline{X}_{\tau_a}\geqslant 0}$ is 
\begin{multline*}
\mathbb{P}_x \PAR{Y_{\tau_a}-a \in dh|~ \underline{X}_{\tau_a}\geqslant 0}\\
 =  \frac{{e^{-\lambda(a,0)(x\vee (h+a)-x\vee a)} \int_0^a  \nu(a-y+dh) R_a^{(0)}(dy)\ind_{h>0}+ \Delta^{(0)}(a) \delta_0(dh)}}{1 - \int_{y\in[0,a]}\int_{h>0} \PAR{1-e^{-\lambda(a,0)h}} 
\nu(x\vee a-y+dh) R_a^{(0)}(dy)}\;,
\end{multline*}
\item the conditional distribution of the maximum reserve level attained before critical  drawdown of size $a$ given the event $\BRA{\underline{X}_{\tau_a}\geqslant 0}$ is 
\begin{multline*}
\mathbb{P}_x\PAR{\overline{X}_{\tau_a} \in dv|~ \underline{X}_{\tau_a}\geqslant 0}\\
=  \frac{\lambda(a,0)e^{-\lambda(a,0)(v-x\vee a)} \PAR{ \int_0^{v-a}\int_0^a\nu(a-y+dh) R_a^{(0)}(dy)+\Delta^{(0)}(a)}}{1 - \int_{y\in[0,a]}\int_{h>0} \PAR{1-e^{-\lambda(a,0)h}} 
\nu(x\vee a-y+dh) R_a^{(0)}(dy)}\ind_{v\geqslant x\vee a} dv\;.
\end{multline*}
\end{enumerate}
\end{pro}

\begin{proof} It is clear from Theorem \ref{main1} that 
\begin{equation}\label{nonruin}
 \mathbb{P}_x(\underline{X}_{\tau_a}\geqslant 0)= \frac{W(x\wedge a)}{W(a)}\PAR{ 1 - \int_{y\in[0,a]}\int_{h>0} \PAR{1-e^{-\lambda(a,0)h}} 
\nu(x\vee a-y+dh) R_a^{(0)}(dy)}.
\end{equation}
  Using Theorem $\ref{main}$ with $u=0$ and $r=q=0$ yields:

\begin{enumerate}

\item  
For $y\in[0,a)$, we have
\begin{align}\label{inter}
\mathbb{P}_x(Y_{\tau_a-} \in dy  ;\underline{X}_{\tau_a}\geqslant 0)&= R_a^{(0)}(dy) 
\frac{W(x\wedge a)}{W(a)} \int_{x\vee a}^\infty F_{0,0,a}(v-(x\vee a)) \int_{h\in(0,v-a]} \nu(a-y+dh) dv
\end{align}
 and for $y=a$, 
$
\mathbb{P}\PAR{Y_{\tau_a-}=a ;\underline{X}_{\tau_a}\geqslant 0}=\frac{W(x\wedge a)}{W(a)}\Delta^{(0)}(a) \int_{x\vee a }^{\infty}F_{0, 0,a}(v-x\vee a) dv.
$ Using Fubini's Theorem and equation \eqref{function} in \eqref{inter} we have

\small
\begin{equation}\label{intersec1}
\mathbb{P}_x\PAR{Y_{\tau_a-} \in dy; \underline{X}_{\tau_a}\geqslant 0 }\\
= \frac{W(x\wedge a)}{W(a)}\SBRA{R_a^{(0)}(dy)   e^{\lambda(a,0)x\vee a}\int_{h>0} e^{-\lambda(a,0)(x\vee (h+a))} 
\nu(a-y+dh) + \Delta^{(0)}(a) \delta_0(dy)}
\end{equation}
\normalsize

The first part of the Theorem is obtained by using \eqref{intersec1} and \eqref{nonruin}.
 
\item For $h>0$, we have
\begin{equation}\label{intersect2}
\mathbb{E}_x\SBRA{\ind_{\{Y_{\tau_a}-a \in dh \}} ;\underline{X}_{\tau_a}\geqslant 0}= \frac{W(x\wedge a)}{W(a)} \int_{x\vee(h+a)}^{\infty}F_{0, 0,a}(v-x\vee a) dv\int_0^a\nu(a-y+dh)R_a^{(0)}(dy),
\end{equation}
and for $h=0$, $
\mathbb{P}\PAR{Y_{\tau_a}=a ;\underline{X}_{\tau_a}\geqslant 0}= \frac{W(x\wedge a)}{W(a)}\int_{x\vee a}^{\infty}F_{0, 0,a}(v-{x\vee a}) dv\Delta^{(0)}(a).$ The proof of the second part of the Theorem is done by applying  the last equation, \eqref{intersect2} and \eqref{nonruin} into the definition of conditional distribution.

\item At the end, for $v\geqslant {x\vee a}$
\begin{equation}\label{intersecttt}
\mathbb{P}_x(\overline{X}_{\tau_a} \in dv ;\underline{X}_{\tau_a}\geqslant 0)=\frac{W(x\wedge a)}{W(a)} F_{0, 0,a}(v-x\vee a) \PAR{ \int_0^{v-a}\int_0^a\nu(a-y+dh) R_a^{(0)}(dy)+\Delta^{(0)}(a)}dv.
\end{equation}
The proof is complete by using \eqref{intersecttt} and \eqref{nonruin}.
\end{enumerate}
\end{proof}

\begin{pro}\label{conditionall}
Consider an insurance risk process $(X_t)_{t\geqslant 0}$ satisfying assumptions of Section \ref{sec:Depletion}  with initial surplus $x> 0$ and let $a>0$ be a fixed critical drawdown size. Then, for $q,r\geqslant 0$, 
 we have
\begin{align}\label{Laplace and nonruin}
&\mathbb{E}_x \left[ e^{-q \tau_a -r\overline{G}_{\tau_a}} ;\underline{X}_{\tau_a}\geqslant 0\right ]\nonumber\\
&=
\frac{W^{(q+r)}(x\wedge a) }{W^{(q+r)}(a)}\frac{\lambda(a,q)}{\lambda(a,q+r)}
\PAR{\int_{y\in[0,a]} \int_{h>0} e^{-\lambda(a,q+r) (x\vee (h+a)-x\vee a)} \nu(a-y+dh) R_a^{(q)}(dy) + \Delta^{(q)}(a)}.
\end{align}

\end{pro}
\begin{proof}
Like in the proof of the previous proposition, the result is obtained by putting $u=0$ in $(15)$ and $(16)$ in Theorem \ref{main} and integrating them with respect to $v$, $h$ and $y$ and switching integrals $dv$ and $dh$.
\end{proof}
We notice that in general on the event $\BRA{\underline{X}_{\tau_a}\geqslant 0}$, $\tau_a -\overline{G}_{\tau_a}$ and $\overline{G}_{\tau_a}$ are no more independent variables (especially when the diffusion coefficient $\sigma$ is positive).

\section{Three Examples of L\'evy Insurance Risk Processes}
\label{sec:examples}

We study in this section three examples of risk process $X$ satisfying the general setting described in Section \ref{sec:Depletion}, starting at an initial surplus $x\geq0$, without Brownian motion part, i.e. $\sigma=0$ in its Laplace exponent (\ref{khinchine}).

Thus the results in Section \ref{sec:general} apply to our problem and they endow us with tools to fully study the depletion problem. Since $\sigma=0$ in the studied models, the set $A_c$ in Theorem \ref{main} is empty and the coefficient $\Delta^{(q)}(a)$ will not appear in the sequel. In this paper, we aim at computing expressions for the distribution of the depletion-related random variables for relevant insurance risk processes. As it turns out, the results in Theorems \ref{main}, \ref{main1}, \ref{main2} and \ref{Laplace_speed} lead to explicit expressions for the distribution of depletion-related random variables when the $q$-scale function of the model has a tractable form. In fact, we will see that a tractable form for the $q$-scale function is inherited by the functions $\lambda$, $F_{p,q,a}$ and $R_a^{(q)}$ defined in (\ref{Lambda}), (\ref{function}) and (\ref{resolvent}) respectively. These functions are the key ingredients in the general expressions of Theorems \ref{main}, \ref{main1}, \ref{main2} and \ref{Laplace_speed}. In this section, we show how there are some interesting examples of insurance models with tractable $q$-scale functions leading to relatively simple expressions for the distributions of depletion random variables. In the following, we will analyze in more detail three models for which we can have an explicit understanding of the depletion problem:
\begin{itemize}
\item the Classical Cramer-Lundberg model with exponential claims,
\item the Gamma risk process,
\item and the Spectrally Negative Stable risk process.
\end{itemize} 

\subsection{Classical Cramer-Lundberg Model with Exponential Claims}


The so-called classical or Cramer-Lundberg model was introduced in \cite{lundberg}. The risk process $X$ is a compound Poisson process starting at $x\geqslant 0$, i.e. 
\begin{equation}
\label{classrisk}
X_t=x+ct-\sum_{i=1}^{N_t}Z_i\;,
\end{equation}
where the number of claims is assumed to follow a Poisson process $(N_t)_{t\geqslant 0}$ with intensity $\lambda$.  which is independent of the positive and \emph{iid} random variables $(Z_n)_{n\geqslant 1}$ representing claim sizes. The loaded premium $c$ is of the form $c= (1+\theta) \lambda \mathbb E[Z_1]$ for some safety loading factor $\theta>0$. The form of the $q$-scale function in this model is relatively simple when the claim sizes are exponentially distributed with mean $1/\mu$. In this case, the L\'evy measure takes the simple form $\nu(dx)=\lambda K(dx)$ where $K$ is the exponential probability measure associated to the claim sizes. In turn, the Laplace exponent in (\ref{khinchine}) becomes,
\begin{equation}\label{laplace_classical}
\psi(s) = c \,s - \lambda [\phi_K(s)-1]\;, \qquad s>0\;,
\end{equation}
where $\phi_K(s)=\frac{\mu}{\mu+s}$ is the Laplace transform of an exponential distribution (see for instance \cite{Kyprianou}). 
In this case, the premium rate is $c=\frac{\lambda(1+\theta)}{\mu}$ where $\theta>0$ is a positive security loading.

A path of a such process is linear by part, so its corresponding drawdown process is quite simple to draw.
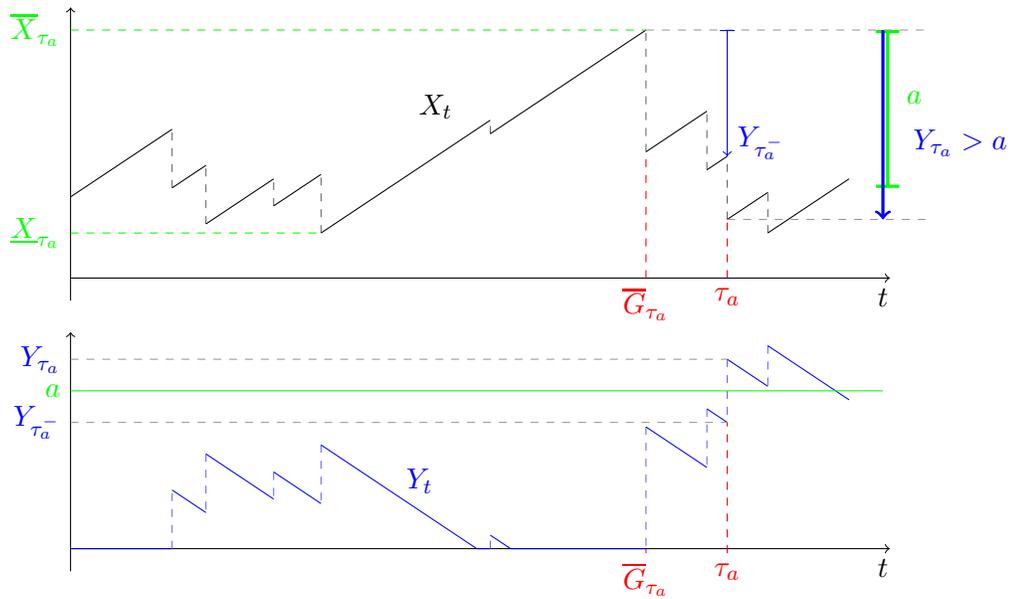
\begin{figure}[h!]\label{fi:general}
\begin{center}
%
%
%
%
%
%

\begin{tikzpicture}[xscale=0.9,yscale=0.6]
\begin{scope}
\draw[->] (0,1) -- (12.1,1);
\draw (12,1) node[below] {$t$};
\draw [->] (0,0.5) -- (0,7);
\draw (5,4.8) node[color=black,right] {$X_t$};

\draw [color=black] (0,2.8) -- (1.5,4.3);
\draw [dashed, color=black!60] (1.5,3) -- (1.5,4.3);

\draw [color=black] (1.5,3) -- (2,3.5);
\draw [dashed, color=black!60] (2,2.2) -- (2,3.5);

\draw [color=black] (2,2.2) -- (3,3.2);
\draw [dashed, color=black!60] (3,2.6) -- (3,3.2);

\draw [color=black] (3,2.6) -- (3.7,3.3);
\draw [dashed, color=black!60] (3.7,2) -- (3.7,3.3);

\draw [color=black] (3.7,2) -- (6.2,4.5);
\draw [dashed, color=black!60] (6.2,4.2) -- (6.2,4.5);

\draw [color=black] (6.2,4.2) -- (8.5,6.5);
\draw [dashed, color=black!60] (8.5,3.8) -- (8.5,6.5);

\draw [color=black] (8.5,3.8) -- (9.4,4.7);
\draw [dashed, color=black!60] (9.4,3.4) -- (9.4,4.7);

\draw [color=black] (9.4,3.4) -- (9.7,3.7);
\draw [dashed, color=black!60] (9.7,2.3) -- (9.7,3.7);

\draw [color=black] (9.7,2.3) -- (10.3,2.9);
\draw [dashed, color=black!60] (10.3,2) -- (10.3,2.9);

\draw [color=black] (10.3,2) -- (11.5,3.2);


\draw [dashed, color=green] (0,6.5) -- (8.5,6.5);
\draw [color=green](0,6.5) node[left] {$\overline{X}_{\tau_a}$};

\draw [dashed, color=red] (8.5,1) -- (8.5,3.8);
\draw [color=red](8.5,1) node[below] {$\overline{G}_{\tau_a}$};

\draw [dashed, color=red] (9.7,1) -- (9.7,2.3);
\draw [color=red](9.7,1) node[below] {$\tau_a$};

\draw[|-|][line width=1.2pt,color=green] (12.07,6.5) -- (12.07,3);
\draw(12.2,5) node[right,color=green] {$a$};

\draw[->][line width=1.2pt,color=blue] (12,6.5) -- (12,2.3);
\draw [color=blue](12.3,4) node[right] {$Y_{\tau_a}>a$};

\draw [dashed,color=black!40] (8.5,6.5) -- (12.7,6.5);
\draw [dashed,color=black!40] (9.7,2.3) -- (12.7,2.3);

\draw[|->,color=blue](9.7,6.5) -- (9.7,3.7);
\draw [color=blue](9.7,4) node[right] {$Y_{\tau_a^-}$};

\draw [dashed, color=green] (0,2) -- (3.7,2);
\draw [color=green](0,2) node[left] {$\underline{X}_{\tau_a}$};
\end{scope}

\begin{scope}[xshift=0cm,yshift=-5cm]
\draw[->] (0,0) -- (12.1,0);
\draw (12,0) node[below] {$t$};
\draw [->] (0,-0.5) -- (0,4.8);
\draw (5.5,1.5) node[color=blue,left] {$Y_t$};

\draw [color=blue] (0,0) -- (1.5,0);
\draw [dashed, color=blue!60] (1.5,0) -- (1.5,1.3);

\draw [color=blue] (1.5,1.3) -- (2,0.8);
\draw [dashed, color=blue!60] (2,2.1) -- (2,0.8);

\draw [color=blue] (2,2.1) -- (3,1.1);
\draw [dashed, color=blue!60] (3,1.7) -- (3,1.1);

\draw [color=blue] (3,1.7) -- (3.7,1);
\draw [dashed, color=blue!60] (3.7,2.3) -- (3.7,1);

\draw [color=blue] (3.7,2.3) -- (6,0);
\draw [color=blue] (6,0) -- (6.2,0);

\draw [dashed, color=blue!60] (6.2,0) -- (6.2,0.3);
\draw [color=blue] (6.2,0.3) -- (6.5,0);

\draw [color=blue] (6.5,0) -- (8.5,0);
\draw [dashed, color=blue!60] (8.5,2.7) -- (8.5,0);

\draw [color=blue] (8.5,2.7) -- (9.4,1.8);
\draw [dashed, color=blue!60] (9.4,3.1) -- (9.4,1.8);

\draw [color=blue] (9.4,3.1) -- (9.7,2.8);
\draw [dashed, color=blue!60] (9.7,4.2) -- (9.7,2.8);

\draw [color=blue] (9.7,4.2) -- (10.3,3.6);
\draw [dashed, color=blue!60] (10.3,4.5) -- (10.3,3.6);

\draw [color=blue] (10.3,4.5) -- (11.5,3.3);
%

\draw [color=red](8.5,-0.1)--(8.5,0);
\draw [color=red](8.5,-0.1) node[below] {$\overline{G}_{\tau_a}$};

\draw [dashed, color=red] (9.7,-0.1) -- (9.7,2.8);
\draw [color=red](9.7,-0.1) node[below] {$\tau_a$};

\draw [dashed, color=black!40] (0,2.8) -- (9.7,2.8);
\draw [color=blue](0,2.8) node[left] {$Y_{\tau_a^-}$};

\draw[-][color=green] (0,3.5) -- (12,3.5);
\draw(0,3.5) node[left,color=green] {$a$};

\draw[-][dashed,color=black!40] (0,4.2) -- (9.7,4.2);
\draw [color=blue](0,4.2) node[left] {$Y_{\tau_a}$};
\end{scope}
\end{tikzpicture}

\caption{A path of a compound Poisson process $X$, the corresponding drawdown process $Y$ and their related depletion quantities.}
\end{center}
\end{figure}

This model has been for long a textbook example for which the distribution of ruin-related quantities can be explicitly computed. This is in fact possible thanks to the tractable form of the $q$-scale function although maybe this was not immediately recognized. It turns out that just like for the ruin problem, this tractable form of the $q$-scale function also allows for an explicit study of the depletion problem. Here, we give the form of the $q$-scale function and study in detail the depletion problem for this particular example. Moreover, we derive explicit expressions for the distributions of depletion random variables of Theorems \ref{main}, \ref{main1}, \ref{main2} and \ref{Laplace_speed}.

The expression for the $q$-scale function in this case is known and is given by 

\begin{equation}
\label{scalefunction}
W^{(q)}(x)=\frac{1}{c^2\PAR{\Phi(q)+\mu}^2-c\lambda\mu }\SBRA{c\PAR{\Phi(q)+\mu}^2\,e^{\Phi(q)x} -\lambda \mu e^{-\PAR{\mu -\frac{\lambda \mu}{c(\Phi(q)+\mu)}} x}}
\end{equation}
with $\Phi(q)= {1\over 2c}\PAR{q+\lambda -c\mu +\sqrt{(q+\lambda-c\mu )^2+4q\mu c}}$. For details see \cite{Kuz_Kyp_Riv}.

Now we also need expressions for the functions $\lambda$ and $R_a^{(q)}$. These are given in the following result.
\begin{pro}\label{pro:L}
Consider the process $ X_t$ in (\ref{classrisk}) where the $Z_i$'s are identically independent exponential random variables with mean $1/\mu$ and let $a>0$ be a critical drawdown size. Then,
\begin{align}\label{prop_Lambda}
 \lambda(a,q)
 &= \Phi(q) +\frac{\lambda \mu \SBRA{c\PAR{\Phi(q)+\mu}^2-\lambda \mu}}{c\PAR{\Phi(q)+\mu}\SBRA{\PAR{\Phi(q)+\mu}^2ce^{\PAR{\Phi(q)+\mu-\frac{\lambda\mu}{\PAR{\Phi(q)+\mu}c}}a}-\lambda \mu}}\;,\\
 \notag
R_a^{(q)}(dy)&={1\over c\lambda(a,q)}\delta_0+\SBRA{{1\over \lambda(a,q)}W'^{(q)}(y)-W^{(q)}(y)}\ind_{y\in(0,a]}dy\;,
\end{align}
with $W^{(q)}$ and $W'^{(q)}$ given by \eqref{scalefunction} and \eqref{derivative} respectively.
\end{pro}
\begin{proof}\noindent
By definition,
$$ \lambda(a,q) = \frac{W_+'^{(q)}(a)}{W^{(q)}(a)} \;. $$
By referring to  equation (\ref{scalefunction}) we can see that $W^{(q)}$ is a derivable function on $(0,\infty)$ so by differentiating equation  (\ref{scalefunction}) we can directly obtain,
\begin{eqnarray}
W'^{(q)}(x) &=&\Phi(q)W^{(q)}(x)+\frac{\lambda \mu}{c\SBRA{\PAR{\Phi(q)+\mu}^2c-\lambda\mu}}\SBRA{\phi(q)+\PAR{\mu -\frac{\lambda \mu}{(\Phi(q)+\mu)c} }}e^{-\PAR{\mu -\frac{\lambda \mu}{(\Phi(q)+\mu)c}} x}\nonumber\\
&=&\Phi(q)W^{(q)}(x)+\frac{\lambda \mu}{c^2\PAR{\Phi(q)+\mu}}e^{-\PAR{\mu -\frac{\lambda \mu}{(\Phi(q)+\mu)c}} x}.
\label{derivative}
\end{eqnarray}
Combining (\ref{scalefunction}) and  (\ref{derivative}) in the definition (\ref{Lambda}) yields the first result.

 As $W^{(q)}$ is an increasing function and has a mass at $x=0$, recall that in this case (see \cite{Kyprianou}), $W^{(q)}(0)=1/c$,  we have  $W^{(q)}(dx)=W'^{(q)}(x) dx + \frac{1}{c} \delta_0(dy)$. Direct substitution into the definition (\ref{resolvent}) of $R_a^{(q)}$ yields the second result.
\end{proof}

The main results regarding depletion-related quantities are given in terms of $W^{(q)}$ and $\lambda(a,q)$ with $q=0$. These expressions take on a more simple form in this case and are given in the following result. 

\begin{pro}\label{remark1}
Consider the process $ X_t$ in (\ref{classrisk}) where the $Z_i$'s are identically independent exponential random variables with mean $1/\mu$ and let $a>0$ be a critical drawdown size. Then,
\begin{align*}
 W(x)&= \frac{\mu}{\lambda(1+\theta)\theta} \PAR{1+\theta-e^{\frac{-\mu \theta }{1+\theta}x}}\;,\\[0.1in]
W'(x)&= \frac{\mu^2}{\lambda(1+\theta)^2}  ~ e^{\frac{-\mu \theta }{1+\theta}x}\;,\\[0.1in]
\lambda(a,0)& = \frac{\mu \theta}{(1+\theta)\left( (1+\theta) e^{\frac{\mu \theta}{1+\theta}a} -1\right)}\;,\\[0.1in]
F_{0,0,a}(y) &=\frac{\mu \theta}{(1+\theta)\left( (1+\theta) e^{\frac{\mu \theta}{1+\theta}a} -1\right)}\exp\PAR{\frac{-\mu  \theta y}{(1+\theta)\left( (1+\theta) e^{\frac{\mu \theta}{1+\theta}a} -1\right)}}\;,\\[0.1in]
R_a^{(0)}(dy)& = \frac{\mu}{\lambda \theta} \PAR{e^{\frac{\mu \theta}{1+\theta}(a-y)}-1}\ind_{y\in(0,a]} dy + \frac{1}{\lambda \theta}\left((1+\theta) e^{\frac{\mu \theta}{1+\theta}a} -1\right) \delta_0(dy)\;.
\end{align*} 
 \end{pro}
\begin{proof}
It is straight forward by setting $q=0$ in Propositions \ref{pro:scalefunction} and \ref{pro:L} with $c=\frac{\lambda(1+\theta)}{\mu}$, and by using definitions (\ref{function})and (\ref{resolvent}). Simply recall that $\Phi (0)=0$. 
\end{proof}  

We now give explicit representations for the distributions in Theorem \ref{main1}, \ref{main2} and \ref{Laplace_speed} as they are specialized to this case.

\medskip

\begin{pro}
Consider an insurance risk process $(X_t)_{t\geqslant 0}$ of the form defined (\ref{classrisk}) with an initial level $x\geqslant 0$ and let $a>0$ be a fixed critical drawdown size. Then, 
\begin{equation}\label{ruin_before_CPE}
\mathbb{P}_x(\underline{X}_{\tau_a}<0)=1-\frac{W(x\wedge a)}{W(a)}\PAR{1-\frac{\lambda(a,0) }{\mu + \lambda(a,0)}e^{-\mu(x\vee a - a)}},
\end{equation}


where $\lambda(a,0)$ and the scale function $W$ are given in Proposition \ref{remark1}. 
\end{pro} 

\begin{proof}
To show this proposition we use the expression of  $\mathbb{P}_x\PAR{\underline{X}_{\tau_a}<0}$ given in Theorem \ref{main1}. But in this model $\sigma =0$ so we have $\Delta^{(0)}(a)=0$ and 

\begin{equation}\label{classic}
\mathbb{P}_x(\underline{X}_{\tau_a}<0)=1-\frac{W(x\wedge a)}{W(a)}+\frac{W(x\wedge a)}{W(a)}\int_{y\in[0,a]}\int_{h>0} \PAR{1-e^{-\lambda(a,0)h}} 
\nu(x\vee a-y+dh) R_a^{(0)}(dy),
\end{equation}

The L\' evy measure for the process $S_t$ is $\nu(dx) = \lambda \mu e^{-\mu x} dx$. So, by replacing this into the interior integral in \eqref{classic}  we have

\begin{align}\notag
\int_{h>0} \PAR{1-e^{-\lambda(a,0)h}} 
\nu(x\vee a-y+dh)&=\int_{h>0} \PAR{1-e^{-\lambda(a,0)h}}  \lambda \mu e^{-\mu(x\vee a -y+ h)} dh\\
&=\lambda e^{-\mu(x\vee a)}\PAR{e^{\mu y}-{\mu\over\mu+\lambda(a,0)} e^{\mu y}}\label{factorization7}.
\end{align}

On the other hand, we have

\begin{equation}\label{equation1}
\lambda \int_0^a e^{\mu y} R_a^{(0)}(dy)=e^{\mu a}.
\end{equation} 

 So by applying  \eqref{equation1} and \eqref{factorization7}  into \eqref{classic} we have
\begin{equation*}
\mathbb{P}_x(\underline{X}_{\tau_a}<0)=1-\frac{W(x\wedge a)}{W(a)}+\frac{W(x\wedge a)\lambda(a,0) e^{-\mu(x\vee a - a)}}{W(a) (\mu + \lambda(a,0))}
\end{equation*}

\end{proof}

%
%
%

 \begin{pro}\label{pro:depletion_CPE}
Consider an insurance risk process $(X_t)_{t\geqslant 0}$ of the form defined (\ref{classrisk}) with an initial level $x\geqslant 0$ and let $a>0$ be a fixed critical drawdown size. Then,
\begin{enumerate}
\item the largest drawdown observed before critical drawdown follows a mixture of a diffusive distribution on $(0,a]$ and the Dirac measure at $0$,
\begin{equation*}
\mathbb{P}_x(Y_{\tau_a-} \in dy )= {\mu\over \theta}\PAR{e^{-{\mu\over 1+\theta}(a-y)}-e^{-\mu(a-y)}}\ind_{y\in(0,a]}dy+{1\over \theta}\PAR{(1+\theta)e^{-{\mu\over 1+\theta}a}-e^{-\mu a}}\delta_0(dy)\;,
\end{equation*}
\item the overshoot over the critical drawdown $Y_{\tau_a}-a$ follows an exponential distribution with mean $1/\mu$,
\begin{equation*} 
\mathbb{P}_x (Y_{\tau_a}-a \in dh) = \mu e^{-\mu h} \ind_{h>0}dh\;.
\end{equation*}
\end{enumerate} 
\end{pro}

\begin{proof} 
We prove this proposition by applying Theorem \ref{main2}. Here $\sigma =0$ and naturally $\Delta^{(0)}(a) =0.$
\begin{enumerate}
\item By plugging $\int_0^{\infty} \nu(a-y+dh) = \lambda e^{-\mu(a-y)}$  into $(\ref{eq:LawY-})$ we have
 
 \begin{equation}\label{factorization16}
\mathbb{P}_x(Y_{\tau_a-} \in dy )= \lambda e^{-\mu(a-y)} R_a^{(0)}(dy).
 \end{equation}
We conclude the first part of the theorem using the expression of $R_a^{(0)}(dy)$ given in Proposition \ref{remark1}.

\item To prove the second part of the theorem we have 
\begin{equation}\label{factorization**}
\nu(a-y+dh)=  \lambda \mu e^{-(a-y+h)\mu} dh .
\end{equation}
\noindent for $h>0$, $~y\in[0,a].$  Substituting $(\ref{factorization**})$ in  $(\ref{factorization17})$  gives

$\mathbb{P}_x (Y_{\tau_a}-a \in dh)= \mu e^{-\mu h} dh.$
\end{enumerate}
\end{proof}


We now provide explicit expressions for joint Laplace transform of $\overline{G}_{\tau_a}$ and $\tau_a$ as well as the Laplace transform of the speed of depletion. Notice that the Laplace transform of $\overline{G}_{\tau_a}$ is a simple function of $\lambda(a,q)$ and it can be computed in a straightforward way using Proposition \ref{lemma_L} and equation (\ref{prop_Lambda}). 

\begin{pro}
Consider an insurance risk process $(X_t)_{t\geqslant 0}$ of the form defined (\ref{classrisk}) with an initial level $x\geqslant 0$ and let $a>0$ be a fixed critical drawdown size. Then, for $q,r\geqslant 0$, 
\begin{enumerate}
\item the bivariate Laplace transform of $\tau_a$ and $\overline{G}_{\tau_a}$ is given by
\begin{equation}\label{factorization19}
\mathbb{E}_x \left[ e^{-q \tau_a -r\overline{G}_{\tau_a}}\right ] =\PAR{\frac{\Phi(q)+\mu}{\mu}}\PAR{\frac{\lambda(a,q)-\Phi(q)}{ \lambda(a,q+r)}}e^{\Phi(q)a}\;,
\end{equation}
with $\Phi(q)=(2c)^{-1} \PAR{q+\lambda -c\mu +\sqrt{(q+\lambda-c\mu )^2+4q\mu c}}$, 
\item the Laplace transform of the speed of depletion $\tau_a -\overline{G}_{\tau_a}$ is given by,
\begin{equation*}
 \mathbb{E}_x\SBRA{e^{-q (\tau_a -\overline{G}_{\tau_a})}} =\PAR{\frac{\Phi(q)+\mu}{\mu}}\PAR{\frac{\lambda(a,q)-\Phi(q)}{ \lambda(a,0)}}e^{\Phi(q)a}\;.
 \end{equation*}
\end{enumerate}
Recall that $\lambda(a,.)$ is given in Proposition \ref{remark1}.
\end{pro}

\begin{proof} \noindent 
As $\sigma =0$ so naturally $\Delta^{(0)}(a) =0.$
\begin{enumerate}
\item  The L\' evy measure of the process $X$ is $\nu(dx) = \lambda \mu e^{-\mu x} dx$ and so $\int_0^{\infty} \nu(a-y+dh) = \lambda e^{-\mu(a-y)}$. Now using Proposition \ref{lemma_L} as it specializes to this case yields, 
\begin{equation}\label{factorization20} 
\mathbb{E}_x \left[ e^{-q \tau_a -r\overline{G}_{\tau_a}}\right ] = \frac{\lambda(a,q) \, \lambda e^{-\mu a} }{\lambda(a, q+r)} \int_{y\in[0,a]} e^{\mu y}R_a^{(q)}(dy) \;. 
\end{equation}

Using now Proposition \ref{pro:L} we can compute the following integral,
\small
\begin{equation}\label{factorization21}
\lambda e^{-\mu a}\int_{y\in[0,a]} e^{\mu
 y}R_a^{(q)}(dy) = {\lambda\over \lambda(a,q)}W^{(q)}(a)-\lambda e^{-\mu a}\PAR{{\mu\over \lambda(a,q)}+1} \int_0^a e^{\mu y} W^{(q)}(y) dy.
\end{equation}
\normalsize
In order to finish the proof it would be sufficient to find an expression for $\int_0^a e^{\mu y} W^{(q)}(dy)$. By using equation (\ref{scalefunction}) in Proposition \ref{pro:scalefunction}, we have
\begin{equation}\label{factorization22}
\int_0^a e^{\mu y} W^{(q)}(y) dy= \frac{\Phi(q)+\mu}{\PAR{\Phi(q)+\mu}^2c-\lambda\mu}\PAR{e^{\PAR{\Phi(q)+\mu}a}-e^{\frac{\lambda\mu}{\PAR{\Phi(q)+\mu}c}a}}\;.
\end{equation}
Now, by combining equations $(\ref{factorization22})$ and $(\ref{factorization21})$ into $(\ref{factorization20})$ we obtain
\[
\mathbb{E}_x \left[ e^{-q \tau_a -r\overline{G}_{\tau_a}}\right ] ={\lambda\over \lambda(a,q+r)}\PAR{W^{(q)}(a)-\PAR{\mu+\lambda(a,q)} \frac{\Phi(q)+\mu}{\PAR{\Phi(q)+\mu}^2c-\lambda\mu}\PAR{e^{\Phi(q)a}-e^{-\PAR{\mu-\frac{\lambda\mu}{\PAR{\Phi(q)+\mu}c}}a}}}
\]
Using the expressions \eqref{scalefunction} and \eqref{prop_Lambda} for $W^{(q)}(a)$ and  $\lambda(a,q)$ respectively, we deduce that,
\begin{align*}
\mathbb{E}_x \left[ e^{-q \tau_a -r\overline{G}_{\tau_a}}\right ]&={\lambda\over \lambda(a,q+r)}\frac{\PAR{(\Phi(q)+\mu)^2c-\lambda\mu}e^{\Phi(q)a}}{c\PAR{(\Phi(q)+\mu)^2ce^{\PAR{\Phi(q)+\mu-{\lambda\mu\over \PAR{\Phi(q)+\mu}c}}a}-\lambda\mu}}\\
&=\frac{\PAR{\Phi(q)+\mu}\PAR{\lambda(a,q)-\Phi(q)}e^{\Phi(q)a}}{\mu \lambda(a,q+r)}.
\end{align*}
This completes the proof.
\item In a similar way, specializing Theorem \ref{Laplace_speed} to this case using the expressions in Propositions \ref{pro:scalefunction} and \ref{pro:L} yields the result.
\end{enumerate}
\end{proof}
In the following proposition we provide the conditional distribution for the depletion quantities given the event $\{\underline{X}_{\tau_a}\geqslant 0\}$. Expressing the joint Laplace transform for $\overline{G}_{\tau_a}$ and $\tau_a$ in the presence of the event $\{\underline{X}_{\tau_a}\geqslant 0\}$
is also of interest  so that for which we discuss in the following proposition.
\begin{pro}\label{conditional cases}
Consider an insurance risk process $(X_t)_{t\geqslant 0}$ of the form defined (\ref{classrisk}) with an initial level $x\geqslant 0$ and let $a>0$ be a fixed critical drawdown size. Then, 
\begin{enumerate}
\item $\mathbb{P}_x\PAR{Y_{\tau_a-} \in dy| ~\underline{X}_{\tau_a}\geqslant 0 } = \lambda e^{-\mu(a-y)} R_a^{(0)}(dy) \ind_{[0,a]}(y),$
\item $\mathbb{P}_x\PAR{Y_{\tau_a}-a \in dh| ~\underline{X}_{\tau_a}\geqslant 0 } = \frac{\mu e^{-\lambda(a,0)(x\vee (h+a)-x\vee a) -\mu h} dh}{1 - \frac{\lambda(a,0)}{\mu + \lambda(a,0)} e^{-\mu(x\vee a -a)}} \ind_{h>0},$
\item $\mathbb{P}_x\PAR{\overline{X}_{\tau_a} \in dv| ~\underline{X}_{\tau_a}\geqslant 0 } = \frac{\lambda(a,0) e^{-\lambda(a,0) x\vee a} \PAR{e^{-\lambda(a,0)v}- e^{-(\lambda(a,0)+\mu) v +\mu a } }}{1 - \frac{\lambda(a,0)}{\mu + \lambda(a,0)} e^{-\mu(x\vee a -a)}} \ind_{v\geqslant x\vee a} dv,$
\item $ \mathbb{E}_x \left[ e^{-q \tau_a -r\overline{G}_{\tau_a}} ;\underline{X}_{\tau_a}\geqslant 0\right ] = \frac{W^{(q+r)}(x\wedge a) }{W^{(q+r)}(a)}\frac{(\mu+\Phi(q) )\PAR{\lambda(a,q) - \Phi(q)}}{ \mu\lambda(a,q+r)} \PAR{1 - \frac{\lambda(a,q+r)}{\mu +\lambda(a,q+r)} e^{-\mu (x\vee a -a)}}.$
\end{enumerate}
\end{pro}
\begin{proof}
To prove this proposition we need to use Proposition \ref{conditional} and \ref{conditionall}. 
\begin{enumerate}
\item To show the first part we compute the expression given in \eqref{intersec1}. In fact by considering different cases between $x$ and $a$ we have 
\begin{equation}\label{intersect3}
\mathbb{P}_x\PAR{Y_{\tau_a-} \in dy; \underline{X}_{\tau_a}\geqslant 0 } = \frac{W(x\wedge a)}{W(a)}\PAR{\lambda e^{-\mu(a-y)} R_a^{(0)}(dy)  \PAR{1 - \frac{\lambda(a,0)}{\mu + \lambda(a,0) }e^{-\mu(x\vee a -a)}}}.
\end{equation}
On the other side, by \eqref{ruin_before_CPE} we have
\begin{equation}\label{nonruin2}
\mathbb{P}_x\PAR{\underline{X}_{\tau_a}\geqslant 0} = \frac{W(x\wedge a)}{W(a)}\PAR{1 - \frac{\lambda(a,0)}{\mu + \lambda(a,0) }e^{-\mu(x\vee a -a)}}
\end{equation}
Applying \eqref{nonruin2} and \eqref{intersect3} to \eqref{intersec1} yields the result in the first part of the theorem.
\item To show this part we also compute the expression given in \eqref{intersect2}. After simplifying the expression we have
\begin{equation}\label{intersect4}
\mathbb{P}_x\PAR{Y_{\tau_a}-a \in dh; \underline{X}_{\tau_a}\geqslant 0 } = \frac{W(x\wedge a)}{W(a)}\PAR{\mu e^{-\lambda(a,0)(x\vee (h+a)-x\vee a) -\mu h}} dh .
\end{equation}
One again applying \eqref{intersect4} and \eqref{nonruin2} to \eqref{intersect2} proves the result in the second part of the theorem.
\item This part can be proven by computing the expression given in \eqref{intersecttt} and using \eqref{nonruin2}. In fact after some simplifications we get,
\begin{equation}\label{intersect5}
\mathbb{P}_x\PAR{\overline{X}_{\tau_a} \in dv; \underline{X}_{\tau_a}\geqslant 0 } = \frac{W(x\wedge a)}{W(a)}\PAR{\lambda(a,0) e^{-\lambda(a,0) x\vee a} \PAR{e^{-\lambda(a,0)v}- e^{-(\lambda(a,0)+\mu) v +\mu a } }} dv .
\end{equation}
To end the proof we just need to replace \eqref{intersect5} and \eqref{nonruin2} to \eqref{intersecttt}.
\item The last part of the theorem can be shown directly by computing the expression in \eqref{Laplace and nonruin}. It is clear from \eqref{factorization19} that 
\begin{equation}\label{Laplaceee}
 \lambda \int_{y\in[0,a]} e^{\mu y}R_a^{(q)}(dy)=\PAR{\frac{\Phi(q)+\mu}{\mu}}\PAR{\frac{\lambda(a,q)-\Phi(q)}{ \lambda(a,q)}}e^{(\Phi(q)+\mu)a}.
\end{equation}
So, the proof is complete if we apply \eqref{Laplaceee} to  the expression in \eqref{Laplace and nonruin} and simplify the expression.
\end{enumerate}
\end{proof}
\begin{rem} 
Under the same assumptions of Theorem \ref{conditional cases}, it can be seen that 
\begin{enumerate}
\item from the first part of the Theorem \ref{conditional cases}, the event $ \{Y_{\tau_a-} \in dy\}$ is independent of the event $ \{ \underline{X}_{\tau_a}\geqslant 0\}$. In fact by recalling \eqref{factorization16} we have
$\mathbb{P}_x\PAR{Y_{\tau_a-} \in dy| ~\underline{X}_{\tau_a}\geqslant 0 } = \lambda e^{-\mu(a-y)} R_a^{(0)}(dy) =\mathbb{P}_x(Y_{\tau_a-} \in dy).$ Thus, knowing $\underline{X}_{\tau_a}$ does not affect on the distribution of the  largest drawdown before critical drawdown, $Y_{\tau_a-}$.
\item  from the second part of the Theorem \ref{conditional cases}, if $x<a$, then the random variable $Y_{\tau_a}-a $ given $\{\underline{X}_{\tau_a}\geqslant 0\}$ follows an exponential distribution with parameter $\mu + \lambda(a,0)$. In other words, 

\begin{equation*}
\mathbb{P}_x\PAR{Y_{\tau_a}-a \in dh| ~\underline{X}_{\tau_a}\geqslant 0 } = (\mu + \lambda(a,0)) 
e^{-(\mu + \lambda(a,0)) h} \ind_{h>0}.
\end{equation*} 

\item from the joint Laplace transform of $(\tau_a,\overline{G}_{\tau_a})$, we deduce that  $\overline{G}_{\tau_a}$ and $\tau_a-\overline{G}_{\tau_a}$ are independent random variables on the event $\BRA{\underline{X}_{\tau_a}\geqslant 0}$. Thus, the Laplace transform of the depletion random variable , $\tau_a-\overline{G}_{\tau_a}$, given the event $\BRA{\underline{X}_{\tau_a}\geqslant 0}$ is
\[
\mathbb{E}_x\SBRA{e^{-q\PAR{\tau_a-\overline{G}_{\tau_a}}} | ~\underline{X}_{\tau_a}\geqslant 0}={\mu+\Phi(q)\over \mu}{\lambda(a,q)-\Phi(q)\over \lambda(a,0)}.
\]
\end{enumerate}
\end{rem}
\subsection{Gamma Risk Process}

The gamma risk model was introduced in \cite{Dufresne_Gerber_Shiu_91} and is defined by
\begin{equation}\label{gamma_risk}
X_t=x+ct-S_t\;,
\end{equation}
where the aggregate claims process $(S_t)_{t\geqslant 0}$ is assumed to follow a gamma subordinator with L\'evy measure 
$$ \nu(dx)= \alpha x^{-1}e^{-\beta x}dx \;, \qquad x>0 \;,$$
where $\alpha, \beta>0$.
The loaded premium $c$ is of the form $c= (1+\theta) \mathbb E[S_1]$ for some safety loading factor $\theta>0$. In turn, the Laplace exponent in (\ref{khinchine}) becomes,
\begin{equation*}
\psi(s) = c \,s - \alpha \ln(1+\frac{s}{\beta})\;. \qquad s>0\;,
\end{equation*}
We refer the reader to \cite{Garrido_Morales} for a discussion of subordinator models in risk theory. 


In this section we are going to provide 
expressions for $ \underline{X}_{\tau_a},\overline{X}_{\tau_a}, Y_{\tau_a-}$ and $ Y_{\tau_a}-a$ associated to the process $X$ given by $X_t = ct - S_t$.\\

To find these expressions we need first to provide $W(x)$ for $X$. Let the process $X$ start at $x\geqslant 0$. Based on the result given in Chapter 8 of \cite{Kyprianou} for survival probability for a spectrally negative L\' evy process we have,
\begin{equation}\label{Survival1}
1-\phi(x)= \left\{\begin{array}{cc}
 \psi_X'(0^+) W(x) &~~~~~~\hbox{if}~~~~ \psi_X'(0^+)>0\\
  0 &  \hbox{Otherwise},
\end{array}\right.
\end{equation}

\noindent where $\phi(x)$ is the probability of ruin and $\psi_X$ is the Laplace exponent for $X$. On the other hand, \cite{Dufresne_Gerber_Shiu_91} gives another
 expression for survival probability for $X_t = ct - S_t$ when $S$ is a gamma subordiantor. That is, 
 \begin{equation}\label{Survival2}
 1-\phi(x)= \frac{\theta}{1+\theta} \sum_{n\geqslant 0} \frac{1}{(1+\theta)^n} M^{*n}(x),
 \end{equation}
 \noindent where $M(x) = \beta \int_0^x \int_{\beta t}^{\infty} u^{-1} e^{-u} du dt = 1+\beta x E_1(\beta x)$. Here $E_1(x) = \int_{x}^{\infty} u^{-1} e^{-u} du $ is the exponential integral function and $M^{*n}(x)= \int_0^x M^{*(n-1)}(x-y) M'(y) dy$ is the nth-fold convolution where $M'(y) = \beta E_1(\beta y)$.
 
 As $\psi_X'(0^+) = c - \psi_S'(0^+) = c - \frac{\alpha}{\beta}$ and $c=\frac{\alpha (1+\theta)}{\beta}$, we have $\psi_X'(0^+) = \frac{\alpha \theta}{\beta} >0$. Now, by equalizing ($\ref{Survival1})$ and ($\ref{Survival2})$ we can get the expression for $W(x)$. More precisely, we have,
 
 \begin{equation}\label{Scale}
 W(x) = \frac{\beta}{(1+\theta) \alpha} \sum_{n\geqslant 0} \frac{1}{(1+\theta)^n} M^{*n}(x). 
 \end{equation}
 
 Taking derivative of $W(x)$ in $(\ref{Scale})$ yields,
 \small
 \begin{equation}
 \label{Dscale}
 W'(x) = \frac{\beta}{(1+\theta) \alpha} \sum_{n\geqslant 0} \frac{1}{(1+\theta)^n} (M^{*n})'(x) = \frac{\beta}{(1+\theta) \alpha} \sum_{n\geqslant 0} \frac{1}{(1+\theta)^n} \int_0^x \frac{\partial M^{*(n-1)}}{\partial x}(x-y)*M'(y) dy.
 \end{equation}
\normalsize
 This yields to the following expression for $\lambda(a,0)$,

\begin{equation}\label{division}
\lambda(a,0) = \frac{W'(a)}{W(a)} = \frac{ \sum_{n\geqslant 0} \frac{1}{(1+\theta)^n} (M^{*n})'(a)}{\sum_{n\geqslant 0} \frac{1}{(1+\theta)^n} M^{*n}(a)}.
\end{equation}

Using $(\ref{Scale})$, $(\ref{Dscale})$ and $(\ref{division})$ can also provide expressions for $F_{0,0,a}(x)$ and $R_a^{(0)}(dy)$.

\medskip
\begin{pro}\label{representation6}
Consider an insurance risk process $(X_t)_{t\geqslant 0}$ of the form defined  \eqref{gamma_risk} with an initial level $x\geqslant 0$ and let $a>0$ be a fixed critical drawdown size. Then, 

\begin{equation*}
 \mathbb{P}_x[\underline{X}_{\tau_a}<0]= 1-\frac{W(x\wedge a)}{W(a)}+\frac{W(x\wedge a) e^{-\beta (x\vee a) }}{W(a)}
 \left(G_{\beta, \beta}(x\vee a) -G_{\beta+ \lambda(a,0), \beta}(x\vee a) \right).
\end{equation*}

where $G_{\gamma, \beta}(v)$, $\lambda(a,0)$ and $W(a)$ are given by \eqref{factorization8'}, $(\ref{division})$ and $(\ref{Scale})$.

\end{pro}

\begin{proof} 

Once again like the procedure we have done in the previous subsection, to show this proposition we need to use Theorem \ref{main1}. Here as $\sigma =0$ so naturally $\Delta^{(0)}(a) =0.$ We just need to compute the integrals in the expression given for $\mathbb{P}_x[\underline{X}_{\tau_a}<0]$
in Theorem \ref{main1}.
The L\' evy measure for the process $S_t$ is $\nu(dx) = \alpha x^{-1} e^{-\beta x} dx$. So, by replacing this into $\int_0^{v-a} \nu(a-y+dh)$ we have
\begin{multline*}
\int_{h>0} \PAR{1-e^{-\lambda(a,0)h}} 
\nu(x\vee a-y+dh)= \alpha\int_{0}^{\infty} \PAR{1-e^{-\lambda(a,0)h}} \PAR{x\vee a-y+h}^{-1} e^{-\beta \PAR{x\vee a-y+h}} dh \\
= \alpha \PAR{E_1\PAR{\beta(x\vee a-y)}-e^{\lambda(a,0)\PAR{x\vee a-y}}E_1\PAR{(\beta+\lambda(a,0))(x\vee a-y)}},
\end{multline*}
\noindent where $E_1$ is the exponential integral function.

Now, define the following function.
\begin{equation}\label{factorization8'}
G_{\gamma, \beta}(v) = \int_0^{\infty} e^{-\gamma h} \int_0^a (v+h-y)^{-1} e^{\beta y} R_a^{(0)}(dy) dh
,
\end{equation}

for $\gamma, v>0$. It is clear that $G_{\beta, \beta}(a) = \frac{e^{\beta a}}{\alpha}$ because of
\begin{equation*}
\alpha \int_0^a E_1(\beta(a-y)) R_a^{(0)}(dy)
= 1.
\end{equation*}
 So by applying numerical methods we can compute the function $G_{\gamma, \beta}$. 

To end the proof it is sufficient to apply \eqref{factorization8'} in $(\ref{ruin_before})$. Thus, we have

\begin{equation*}
\mathbb{P}_x(\underline{X}_{\tau_a}<0)= 1-\frac{W(x\wedge a)}{W(a)}+\frac{W(x\wedge a) e^{-\beta x\vee a }}{W(a)}
 \left(G_{\beta, \beta}(x\vee a) -G_{\beta+ \lambda(a,0), \beta}(x\vee a) \right).
\end{equation*}
%
%
%
\end{proof}

Now, we are going to provide representations for distributions of each of the random variables  $Y_{\tau_a-}$ and $ Y_{\tau_a}-a$. 

\begin{pro}\label{representation7}
Consider an insurance risk process $(X_t)_{t\geqslant 0}$ of the form defined  \eqref{gamma_risk} with an initial level $x\geqslant 0$ and let $a>0$ be a fixed critical drawdown size. Then,
\begin{enumerate}
\item the distribution of $Y_{\tau_a-}$, the largest drawdown observed before the critical drawdown of size a, is: 
\begin{equation*}
 \mathbb{E}_x[\ind_{\{Y_{\tau_a-} \in dy \}}]= \alpha E_1(\beta (a-y)) R_a^{(0)}(dy),
 \end{equation*}
 \item the overshoot of critical drawdown over level $a$ is:
\begin{equation*}
\mathbb{E}_x[\ind_{\{Y_{\tau_a}-a \in dh \}}] = \alpha e^{-\beta(a+h)} \int_0^a (v+h-y)^{-1} e^{\beta y} R_a^{(0)}(dy) dh.
\end{equation*} 
\end{enumerate}
\end{pro}

\begin{proof} 
As $\sigma =0$ thus $\Delta^{(0)}(a) =0.$
\begin{enumerate}
\item It is clear that $\int_0^{\infty} \nu(a-y+dh) = \alpha E_1(\beta (a-y)).$ By plugging it into $(\ref{eq:LawY-})$ we have
 
 \begin{equation*}
 \mathbb{E}_x[\ind_{\{Y_{\tau_a-} \in dy \}}]= \alpha E_1(\beta (a-y)) R_a^{(0)}(dy).
 \end{equation*}
\item To prove the second part of the theorem we have 
\begin{equation}\label{factorization***}
\nu(a-y+dh)=  \alpha e^{-\beta (a-y+h)} (a-y+h)^{-1} dh .
\end{equation}
\noindent for $h>0$~, $ y\in[0,a].$  Substituting $(\ref{factorization***})$ in  $(\ref{factorization17})$   gives

\begin{equation*}
\mathbb{E}_x[\ind_{\{Y_{\tau_a}-a \in dh \}}] = \alpha e^{-\beta(a+h)} \int_0^a (v+h-y)^{-1} e^{\beta y} R_a^{(0)}(dy) dh.
\end{equation*}    
\end{enumerate}
\end{proof}

\subsection{ Spectrally Negative Stable process}
\label{sec: Stable}
In this section we are studying the running minimum at the first-passage time over a level $a$ of the drawdown process $Y$. In fact, we study the probability which the running minimum at the first-passage time goes below $0$ when $ X_t$ is a spectrally negative stable process with stability parameter $\alpha \in (1,2)$. Furthermore, we give representations for distributions of each of the stochastic processes $\overline{X}_{\tau_a}, Y_{\tau_a-}$ and $ Y_{\tau_a}-a$ associated to the main process $X$ given above.

Let $ X_t$ be a spectrally negative stable process with stability parameter $\alpha \in (1,2)$ and Laplace exponent $\phi(s) = s^{\alpha}$ for $s>0$. Moreover, the L\'evy measure in \eqref{khinchine} is $\nu(dx)= \frac{\beta}{x^{1+\alpha}} dx$ for $x, \beta>0$. It can be seen (see for example \cite{KyRi, Pistorius}) that \begin{equation}\label{stable1}
W^{(q)}(x) = \alpha x^{\alpha -1} E'_{\alpha, 1}(q x^{\alpha}),
\end{equation}
 for $x,q\geqslant 0$ where $E_{\alpha, 1}(x) = \sum_{k\geqslant 0} \frac{z^k}{\Gamma(1+\alpha k)}$ is the Mittag-Leffler function and $\Gamma$ is the gamma function.
 \begin{pro}\label{pro:LL}
Let $ X_t$ be a stable process with stability parameter $\alpha \in (1,2)$  and let $a>0$ be a critical drawdown size. Then,
\begin{align}
\label{stable2}
 \lambda(a,q)& = \frac{\alpha-1}{a} + q\alpha a^{\alpha -1} \frac{E''_{\alpha, 1}(q a^{\alpha})}{E'_{\alpha, 1}(q a^{\alpha})}\;,
\\[0.1in] \label{stable3}
R_a^{(q)}(dy) &= \SBRA{\alpha y^{\alpha-2} E'_{\alpha, 1}(q y^{\alpha})\left(\frac{\alpha-1}{\lambda(a,q)} - y\right) + \frac{q\alpha^2 y^{2\alpha-2}}{\lambda(a,q)} E''_{\alpha, 1}(qy^{\alpha})} dy\;.
\end{align}
\end{pro}
 
 \begin{proof}
 Taking derivative of \eqref{stable1} with respect to $x$ and substitute it in  $\lambda(a,q)$ given by \eqref{Lambda} (in $R_a^{(q)}$ given by \eqref{resolvent} respectively) yields \eqref{stable2}(\eqref{stable3} respectively).
 \end{proof}
  As a particular case of Proposition \ref{pro:LL}, we have
  \begin{equation}
 \lambda(a,0) = \frac{\alpha-1}{a}\quad\text{and}\quad
\label{stable5}
 R_a^{(0)}(dy) = (\alpha ay^{\alpha-2} - \alpha y^{\alpha-1}) dy.
 \end{equation}

\begin{pro}\label{stable6}
Consider a spectrally negative stable process $(X_t)_{t\geqslant 0}$ with stability parameter $\alpha \in (1,2)$ with an initial surplus $x\geqslant 0$ and let $a>0$ be a fixed critical drawdown size. Then

\begin{equation*}
 \mathbb{P}_x(\underline{X}_{\tau_a}<0)=
 1-\PAR{\frac{x\wedge a}{a}}^{\alpha-1}+ \beta \PAR{\frac{x\wedge a}{a}}^{\alpha-1}\PAR{\frac{H_{0,0,0}(x\vee a)}{\alpha}- H_{\lambda(a,0),\lambda(a,0),0}(x\vee a)}
\end{equation*}

where $\lambda(a,0)$ is given by $(\ref{stable2})$ and $g_{a,\alpha,0}(v)$ is defined by \eqref{stable11}.
\end{pro}

 \begin{proof}
 To show this proposition we need to use Theorem \ref{main1} one more time. The L\' evy measure for the process $-X_t$ is $\nu(dh) = \frac{\beta}{h^{1+\alpha}} dh$ for $\beta>0$. 
 So, by replacing this into $\int_{h>0} \PAR{1-e^{-\lambda(a,0)h}} 
\nu(x\vee a-y+dh)$ we have

\begin{align*}
\int_{h>0} \PAR{1-e^{-\lambda(a,0)h}} 
\nu(x\vee a-y+dh) &=  \int_{h>x\vee a-y}\PAR{1-e^{\lambda(a,0)\PAR{x\vee a-y}}e^{-\lambda(a,0)h}} \frac{\beta}{h^{1+\alpha}} 
dh\\
 &= \frac{\beta}{\alpha} \left[ \frac{1}{(x\vee a-y)^{\alpha}}-e^{\lambda(a,0)\PAR{x\vee a-y}} \int_{h>x\vee a-y} \frac{\alpha e^{-\lambda(a,0)h} }{h^{1+\alpha}}
dh\right].
\end{align*}

%

Now define 
\begin{equation}\label{stable11}
H_{\gamma_1,\gamma_2,q}(v) = \int_0^{\infty} e^{-\gamma_1 h}\int_0^a \frac{e^{-\gamma_2 y}R_a^{(q)}(dy)}{(v-y+h)^{\alpha+1}} dh.
\end{equation}
We can use the numerical methods to compute $H_{\gamma_1,\gamma_2,q}(v) $ for a given value $v$. For the special case $\gamma_1 =\gamma_2=0$, $v=a$ and $q=0$ we have 
\begin{equation}\label{stable12}
H_{0,0,0}(a) = \int_0^a \frac{R_a^{(0)}(dy)}{(a-y)^{\alpha}} = \frac{\alpha}{\beta}, 
\end{equation}
 because of
\[
\int_0^a\int_0^\infty \frac{\beta}{\PAR{a-y+h}^{1+\alpha}} R_a^{(0)}(dy)=1.
\]
To end the proof it is sufficient to replace \eqref{stable11} and $(\ref{stable12})$  into $(\ref{ruin_before})$.

 Therefore, after some simplifications we get

\small
\begin{equation*}
 \mathbb{P}_x(\underline{X}_{\tau_a}<0)=
 1-\PAR{\frac{x\wedge a}{a}}^{\alpha-1}+ \beta \PAR{\frac{x\wedge a}{a}}^{\alpha-1}\PAR{\frac{H_{0,0,0}(x\vee a)}{\alpha}- H_{\lambda(a,0),\lambda(a,0),0}(x\vee a)}
\end{equation*}
\normalsize
 \end{proof}

\begin{pro}\label{stable15}
Consider a spectrally negative stable process $(X_t)_{t\geqslant 0}$ with stability parameter $\alpha \in (1,2)$ with an initial surplus $x\geqslant 0$ and let $a>0$ be a fixed critical drawdown size.  Then,
\begin{enumerate}
\item the distribution of $Y_{\tau_a-}$, the largest drawdown observed before the critical drawdown of size a, is: 
\begin{equation*}
 \mathbb{E}_x[\ind_{\{Y_{\tau_a-} \in dy \}}]=  \frac{\beta y^{\alpha-2}}{ (a-y)^{\alpha-1}} dy,
 \end{equation*}
 \item the overshoot of critical drawdown over level $a$ is:
\begin{equation*}
\mathbb{E}_x[\ind_{\{Y_{\tau_a}-a \in dh \}}] = \beta \int_0^a \frac{R_a^{(q)}(dy)}{(a-y+h)^{\alpha+1}} dh.
\end{equation*}
\end{enumerate}
\end{pro}

\begin{proof} 
 As $\sigma =0$ thus $\Delta^{(0)}(a) =0.$
\begin{enumerate}
\item It can be easily shown that  $\int_0^{\infty} \nu(a-y+dh) = \frac{\beta}{\alpha (a-y)^{\alpha}}.$ By plugging it into $(\ref{eq:LawY-})$ we get
 
 \begin{equation}\label{stable18}
 \mathbb{E}_x[\ind_{\{Y_{\tau_a-} \in dy \}}]= \frac{\beta}{\alpha (a-y)^{\alpha}} R_a^{(0)}(dy).
 \end{equation}
 The proof of first part is done by replacing \eqref{stable5} into \eqref{stable18}.
\item To prove the second part of the theorem we have 
\begin{equation}\label{stable16}
\nu(a-y+dh)=  \frac{\beta}{(a-y+h)^{\alpha+1}} dh .
\end{equation}
  Substituting $(\ref{stable16})$ in  $(\ref{factorization17})$  gives

\begin{equation*}
\mathbb{E}_x[\ind_{\{Y_{\tau_a}-a \in dh \}}] = \beta \int_0^a \frac{R_a^{(q)}(dy)}{(a-y+h)^{\alpha+1}} dh.
\end{equation*}    
\end{enumerate}
\end{proof}

 In the sequel of this subsection  we are going to provide the joint Laplace transform of $\tau_a$ and $\overline{G}_{\tau_a}$. In fact, we are seeking the Laplace transform of the depletion random variable, $\tau_a -\overline{G}_{\tau_a}$. 

\begin{pro}\label{stable20}
Consider a spectrally negative stable process $(X_t)_{t\geqslant 0}$ with stability parameter $\alpha \in (1,2)$ with an initial surplus $x\geqslant 0$ and let $a>0$ be a fixed critical drawdown size. 

\begin{enumerate}
\item the bivariate Laplace transform of $\tau_a$ and $\overline{G}_{\tau_a}$ is given by
 \begin{equation*}
\mathbb{E}_x\SBRA{e^{-q \tau_a - r \overline{G}_{\tau_a}}} =\frac{\beta\lambda(a,q)}{\alpha \lambda(a,q+r)}  H_{0, 0,q}(a),
\end{equation*}
where $H_{0, 0,q}(a)$ is defined by  \eqref{stable11} and $\lambda(a,q)$ is given by \eqref{stable2}

\item the Laplace transform of the speed of depletion $\tau_a -\overline{G}_{\tau_a}$ is given by,
\begin{equation*}
 \mathbb{E}_x\SBRA{e^{-q (\tau_a -\overline{G}_{\tau_a})}} =\frac{\beta\lambda(a,q)}{\alpha\lambda(a, 0)} H_{0, 0,q}(a)\;.
 \end{equation*}
\end{enumerate}
\end{pro}

\begin{proof} \noindent 
 As $\sigma =0$ thus $\Delta^{(0)}(a) =0$ in Proposition \ref{lemma_L} and Theorem \ref{Laplace_speed}.
\begin{enumerate}
\item  The L\' evy measure of the process $-X$ is $\frac{\beta}{x^{\alpha+1}}$ and so $\int_0^{\infty} \nu(a-y+dh) = \frac{\beta}{\alpha(a-y)^{\alpha}}.$  
Now using Proposition \ref{lemma_L} as it specializes to this case yields, 
\begin{equation*}
\mathbb{E}_x \left[ e^{-q \tau_a -r\overline{G}_{\tau_a}}\right ] = \frac{\beta\lambda(a,q)}{\alpha\lambda(a, q+r)} \int_0^a \frac{ R_a^{(q)}(dy)}{(a-y)^{\alpha}} = \frac{\beta\lambda(a,q)}{\alpha\lambda(a, q+r)} H_{0, 0,q}(a).
\end{equation*}
This ends the proof of the first part.
\item In a similar way, specializing Theorem \ref{Laplace_speed} to this case using the expressions in $\eqref{stable2}$ and $\eqref{stable3}$ yields the result.
\end{enumerate}
\end{proof}


\bibliographystyle{abbrv}
\bibliography{literature}

\begin{thebibliography}{10}

\bibitem{AKP}
F.~Avram, A.~E. Kyprianou, and M.~R. Pistorius.
\newblock Exit problems for spectrally negative {L}\'evy processes and
  applications to ({C}anadized) {R}ussian options.
\newblock {\em Ann. Appl. Probab.}, 14(1):215--238, 2004.

\bibitem{Bertoin}
{\relax J}.~Bertoin.
\newblock {\em L\'evy Processes}, volume 121 of {\em Cambridge Tracts in
  Mathematics}.
\newblock Cambridge University Press, 1996.

\bibitem{BK2010}
E.~Biffis and {\relax A. E}.~Kyprianou.
\newblock A note on scale functions and the time value of ruin for {L}\'evy
  risk processes.
\newblock {\em Insurance:Mathematics and Economics}, 46:85 -- 91, 2010.

\bibitem{BM2010}
E.~Biffis and M.~Morales.
\newblock On a generalization of the {G}erber-{S}hiu function to path-dependent
  penalties.
\newblock {\em Insurance:Mathematics and Economics}, 46:92 -- 97, 2010.

\bibitem{Dufresne_Gerber_Shiu_91}
D.~Dufresne, H.~Gerber, and E.~Shiu.
\newblock Risk theory with the gamma process.
\newblock {\em ASTIN Bulletin}, 21(2), 1991.

\bibitem{feller}
W.~Feller.
\newblock {\em An introduction to probability theory and its applications.
  {V}ol. {I}}.
\newblock John Wiley and Sons, Inc., New York, 1957.
\newblock 2nd ed.

\bibitem{Furrer98}
H.~Furrer.
\newblock Risk processes perturbed by a $\alpha$-stable {L}\'evy motion.
\newblock {\em Scandinavian Actuarial Journal}, (1):59 -- 74, 1998.

\bibitem{Garrido_Morales}
J.~Garrido and M.~Morales.
\newblock On the expected discounted penalty function for {L}\'evy risk
  processes.
\newblock {\em North American Actuarial Journal}, 10(4):196--217, 2006.

\bibitem{Gerber98a}
{\relax H.U}.~Gerber and {\relax E.S.W}.~Shiu.
\newblock On the time value of ruin.
\newblock {\em North American Actuarial Journal}, 2(1):48--78, 1998.

\bibitem{Huzak}
M.~Huzak, M.~Perman, H.~Sikic, and Z.~Vondracek.
\newblock Ruin probabilities and decompositions for general perturbed risk
  processes.
\newblock {\em Ann. Appl. Probab.}, 14(3):1378--1397, 2004.

\bibitem{Kuz_Kyp_Riv}
A.~Kuznetsov, {\relax A.E}.~Kyprianou, and {\relax J.C}.~Rivero.
\newblock The theory of scale functions for spectrally negative lévy processes.
\newblock {\em Unknown Journal}, 2061, 2012.

\bibitem{KM}
{\relax A}.~Kuznetsov and {\relax M}.~Morales.
\newblock Computing the finite-time expected discounted penalty function for a
  family of l\'evy risk processes.
\newblock {\em Scandinavian Actuarial Journal}, 2011.

\bibitem{Kyprianou}
{\relax A.E}.~Kyprianou.
\newblock {\em Introductory Lectures on Fluctuations of {L}\'evy Processes with
  Applications}.
\newblock Springer, 2006.

\bibitem{KyRi}
{\relax A.E}.~Kyprianou and {\relax V. M}.~Rivero.
\newblock Special, conjugate and complete scale functions for spectrally
  negative {L}\'evy processes.
\newblock {\em Electronic Journal of Probability}, 13:1672--1701, 2008.

\bibitem{lundberg}
F.~Lundberg.
\newblock Approximerad framställning av sannolikhetsfunktionen. aterförsäkring
  av kollektivrisker.
\newblock {\em Akad. Afhandling. Almqvist. och Wiksell}, 1903.

\bibitem{Pistorius}
{\relax A}.~Mijatovic and {\relax M}.~Pistorius.
\newblock On the drawdown of completely asymmetric levy processes.
\newblock {\em Stochastic Processes and Their Applications}, 22:3812--3836,
  2012.

\bibitem{Morales}
M.~Morales and W.~Schoutens.
\newblock A risk model driven by {L}\'evy processes.
\newblock {\em Appl. Stoch. Models Bus. Ind.}, 19:147--167, 2003.

\bibitem{pistorius04}
M.~R. Pistorius.
\newblock On exit and ergodicity of the spectrally one-sided {L}\'evy process
  reflected at its infimum.
\newblock {\em J. Theoret. Probab.}, 17(1):183--220, 2004.

\bibitem{Z_H}
{\relax H}.~Zhang and {\relax O}.~Hadjiliadis.
\newblock Drawdowns and the speed of market crash.
\newblock {\em Methodology and Computing in Applied Probability},
  14(3):739--752, 2012.

\end{thebibliography}
\end{document}